\documentclass[reprint,nofootinbib,preprintnumbers,superscriptaddress,aps,prd]{revtex4-2}
\usepackage{graphicx}
\usepackage{mathtools,amssymb} 
\usepackage{bm} 
\usepackage{xcolor}
\usepackage{dsfont} 
\usepackage{slashed} 
\usepackage{hyperref}

\newcommand \beq{\begin{eqnarray}}
\newcommand \eeq{\end{eqnarray}}
\renewcommand{\L}{{\cal L}}

\begin{document}
\unitlength=1mm
\allowdisplaybreaks

\title{Interplay between the chiral and deconfinement transitions\texorpdfstring{\\}{} from a Curci-Ferrari-based Polyakov loop potential}

\author{V. Tomas \surname{Mari Surkau}}
\affiliation{Centre de Physique Th\'eorique, CNRS, Ecole polytechnique, IP Paris, F-91128 Palaiseau, France.}

\author{Urko Reinosa}
\affiliation{Centre de Physique Th\'eorique, CNRS, Ecole polytechnique, IP Paris, F-91128 Palaiseau, France.}

\begin{abstract}
   We couple the two-flavor Nambu--Jona-Lasinio model to a gluonic background corresponding to the gauge-field expectation value in the center-symmetric Landau gauge. Low-energy features in this gauge are captured by a center-symmetric extension of the Curci-Ferrari model and provide a good grasp on key aspects of the confinement/deconfinement transition. Within this framework, we can investigate the interplay between the chiral and deconfinement transitions. Compared to other approaches based on multi-parameter Ans\"atze of the Polyakov loop potential fixed from comparison to finite-temperature lattice data, the modeling of the glue sector in the present set-up depends on only one phenomenological parameter that can be fixed by comparison to lattice data in the vacuum. We detail the structure of the phase diagram, with special emphasis on the finite density axis, and compute thermodynamical observables relevant for applications. We also highlight the properties of the recently introduced net quark number response of the medium as a sensible probe of the phases of QCD, in particular as a tool to disambiguate the nature of certain regions of the phase diagram where the use of the Polyakov loops could lead to misinterpretations. Finally, we critically assess the sensitivity of our results to the various parameters, both in the glue sector and in the chiral sector.
\end{abstract}

\date{\today}

\maketitle

\section{Introduction}
Strongly interacting matter as described by Quantum Chromodynamics (QCD) is a key ingredient in the behavior of various physical systems, including the evolution of the early universe and the cores and mergers of neutron stars, making the qualitative and quantitative understanding of its phase structure a crucial but still unresolved task. Thus, for quite some time already, the QCD phase diagram has been at the center of intense theoretical and experimental efforts, see e.g. Ref.~\cite{Borsanyi:2025ttb} or Chapter 7 in Ref.~\cite{Gross:2022hyw} for recent reviews.

The applicability of otherwise powerful perturbative QCD methods based on a weak-coupling expansion \cite{Ghiglieri:2020dpq, Haque:2014rua} is spoiled by the large value of the interaction strength at the temperature or chemical potential scales of physical interest, where the main features of the phase diagram lie. 
The only other \textit{ab initio} tool to study the phase diagram where errors can be reduced systematically is lattice QCD \cite{Aarts:2023vsf}. It relies on Monte-Carlo importance sampling of the Euclidean QCD action on a discretized spacetime, with the errors controlled by the lattice spacing and statistics. At vanishing chemical potential, this approach has lead to the evidence that QCD has a continuous crossover from a phase described by a Hadron Resonance Gas model at low temperatures to a high temperature quark and gluon (plasma) phase described by thermal perturbation theory \cite{Borsanyi:2013EoS, HotQCD:2014kol}. 
Unfortunately, at finite chemical potentials, i.e., in systems with finite net baryon/quark densities, the Euclidean QCD action is no longer real\footnote{More precisely, it is the effective action resulting from the integration of the quark degrees of freedom that becomes non real.}, leading to a non-positive definite functional integral. This feature, known as the sign problem, prevents the random sampling at the base of lattice simulations and calls for the development of other techniques to investigate the phase diagram.

There are powerful approaches based on first-principle functional methods like Dyson-Schwinger equations \cite{Fischer:2018sdj}, or the Functional Renormalization Group \cite{Fu:2022gou}, which formulate QCD as an infinite set of coupled equations. However, these have to be truncated to make solutions possible, and there is \textit{a priori} no controlled way of doing this. While recently there has been a lot of progress in performing consistent truncations and aiming for converged results, see e.g. Refs.~\cite{Ihssen:2024miv, Huber:2025kwy}, the systems of equations quickly increase in complexity, and solutions become computationally expensive.

An alternative strategy is to reduce the problem to the key symmetries and mechanisms believed to play a role and study the behavior of QCD from effective models. They can provide intuitive pictures, but quantitative precision is often hard to justify, and omitted effects limit their range of applicability. Regardless of their shortcomings, they are usually the best avenue for first explorative studies and to build qualitative understanding. Key features of QCD at finite temperature and density are governed by two approximate symmetries that become exact in the opposite limits of vanishing or infinite quark masses. The confinement/deconfinement transition associated with center symmetry is of the first-order type for very massive quarks or in pure gauge theory \cite{Fromm:2011qi, Fukugita:1989yw, Iwasaki:1992ik, Boyd:1996bx}. As for the chiral transition, although initially thought to be first order in the chiral limit \cite{Pisarski:1983ms}, it could in fact correspond to second order type transition according to recent studies \cite{HotQCD:2019xnw, Braun:2020ada, Cuteri:2021ikv, Bernhardt:2023hpr}, or at least to a weakly first-order transition \cite{Fejos:2024bgl}. In QCD, both symmetries are explicitly broken by the finite quark masses, and instead one observes a crossover of their corresponding order parameters, the chiral condensate and the Polyakov loop \cite{Aoki:2006we, Borsanyi:2010ordPar, HotQCD:2018pds}.

Popular models to study spontaneous chiral symmetry breaking include the quark-based Nambu--Jona-Lasinio (NJL) model \cite{Nambu1961DynamicalI, Nambu1961DynamicalII, Klevansky:1992qe, Hatsuda:1994pi, Buballa:2003qv}, and its bosonized extension, the linear sigma or quark-meson (QM) model \cite{Jungnickel:1995fp, Scavenius:2000qd, Gies:2002hq, Tetradis:2003qa, Schaefer:2008hk}. At finite temperatures, they capture chiral symmetry restoration, but lack the aspect of quark confinement related to the center symmetry, or any sort of gluon dynamics. A way to address this issue is by coupling the chiral models to an effective potential for the order parameter associated with center symmetry, leading to the so-called Polyakov-loop extended models (PNJL, PQM), where the chiral and deconfinement crossovers can be studied simultaneously \cite{Fukushima:2003fw, Ratti:2005jh, Roessner:2006xn, Fukushima:2008wg, Fu:2007xc, Schaefer:2007pw, Mao:2009aq}.

Adding confining effects is clearly beneficial, but, since no exact form of the Polyakov loop thermodynamic potential is known, it comes at the cost of even more modeling in approaches that already rely on several parameters to describe the chiral sector. The common tactic is to fit data from lattice Yang-Mills (YM) simulations using a phenomenological, multi-parameter Ansatz, for which various popular parametrizations exist (with anywhere from two to 19 parameters) \cite{Fukushima:2003fw, Ratti:2005jh, Roessner:2006xn, Fukushima:2008wg, Lo:2013hla}. While practical, these do not capture the extent of non-Abelian gauge dynamics, with increased accuracy requiring more parameters of often obscured physical interpretation, and lack any information about the backreactions between gluons and quarks. The latter point can be partially circumvented by adjusting the parameter that sets the critical temperature scale manually, motivated by perturbative Renormalization Group arguments \cite{Schaefer:2007pw, Haas:2013qwp, Herbst:2013ail}. Still, it would clearly be desirable to have a framework in which the gauge sector could be treated more systematically, without requiring as many parameters.

We propose the minimal Curci-Ferrari (CF) model as such a possibility, which adds only a single parameter to the gauge sector and has been shown to successfully capture many aspects of YM theory and QCD down to infrared scales, see e.g. Ref.~\cite{Pelaez:2021tpq} for a review. The model is built by adding a gluon mass scale to the Landau-gauge-fixed YM Lagrangian, and is renormalizable, so amenable to usual quantum field theoretical treatments \cite{Curci:1976bt}. It is motivated by and reproduces the massive, decoupling behavior of the gluon propagator found in lattice simulations of YM theory and QCD \cite{Bogolubsky:2009lan, Bonnet:2000kw, Bonnet:2001glu, Bowman:2004jm, Bowman:2007du, Cucchieri:2008cir, Iritani:2009mp, Maas:2011se, Oliveira:2012eh}, which has recently been shown to appear through the Schwinger mechanism in advanced truncations of functional YM theory as well \cite{Aguilar:2023mdv, Ferreira:2024czk, Ferreira:2025anh}. A gluon mass gap more generally seems related to solving the infrared Gribov ambiguity problem of the Faddeev-Popov gauge fixing procedure \cite{Gribov:1977wm, Singer:1978dk, Maas:2011se, Serreau:2012cg}.

The CF model features Landau pole-free renormalization group trajectories, with the running coupling in the gauge sector remaining moderately small and even decreasing in the infrared \cite{Tissier:2011ey}, which aligns with observations from lattice simulations \cite{Sternbeck:2005tk, Duarte:2016iko, Bogolubsky:2009lan}. This explains how after incorporating the non-perturbatively generated mass scale, obtained by fitting the perturbative expressions for the CF propagators to corresponding lattice YM or QCD data, many aspects of YM theory and QCD have been described within a relatively simple, weakly coupled expansion \cite{Pelaez:2013cpa, Pelaez:2014mxa, Pelaez:2015tba, Pelaez:2024mtq}, with higher loop contributions giving systematic, small corrections towards a description of gluodynamics \cite{Gracey:2019xom, Barrios:2020ubx, Barrios:2022hzr}. Fitting just the pure YM propagator data results in the equivalent of the typical Polyakov loop potentials fixed from the gauge sector, but notably without requiring any finite temperature input, making all finite temperature results genuine predictions of the model. Additionally, the effects of quarks (with realistic masses) on the vacuum gluon propagator have been taken into account self-consistently in the CF model through a double expansion in the pure gauge coupling and the inverse of the number of colors \cite{Pelaez:2020ups}, allowing a gluon mass fit to full QCD correlation functions, and consequent implementation at finite temperature. 

Finally, simple one-loop approximations of the CF thermodynamic potential in a dynamically fixed background gauge\footnote{Background (Landau-DeWitt) gauges represent a convenient way to study the spontaneous breaking of center symmetry in continuum approaches, as the regular Landau gauge-fixed action would explicitly break the center symmetry \cite{DeWitt:1967ub, Braun:2007bx, MariSurkau:2024zfb}.} reproduce the confinement-deconfinement transition of YM theory and heavy quark QCD, using the background as an order parameter \cite{Reinosa:2014ooa, Reinosa:2015oua}. Two-loop corrections show quantitative improvements towards lattice benchmarks \cite{Reinosa:2015gxn, Maelger:2017amh}. Recently, a new center-symmetric way of fixing the background has allowed the calculation of the thermodynamic potential from a more consistent effective action, where the gauge field correlation functions become order parameters for the center symmetry, as supported by lattice simulations \cite{vanEgmond:2021jyx, vanEgmond:2023lfu, vanEgmond:2024ljf, vanEgmond:2025zxf}. This produces quantitatively much more accurate results already at one-loop order, which are stable against renormalization scheme changes and renormalization scale variations \cite{MariSurkau:2024zfb, MariSurkau:2025dfo}. It thus provides a phenomenologically simple, yet efficient gauge/Polyakov loop potential, ready to be implemented with chiral models.

In this work, we provide the first application of this Polyakov loop potential to QCD thermodynamics with light quarks. To describe chiral symmetry breaking, we employ the simple NJL model, which we introduce together with the center-symmetric CF model in Sec.~\ref{sec: framework}. In Sec.~\ref{sec: phase dia}, we study the phase diagram from the usual order parameters, chiral condensate and Polyakov loops, comparing to lattice QCD results and showing the appearance of a critical endpoint (CEP). We discuss the sensitivity of those results to the parameters in the chiral and glue sectors in App.~\ref{app: params}. In Sec.~\ref{sec:T0}, we focus on the $\mu$-axis of the phase diagram corresponding to the $\smash{T\to 0}$ limit, with special emphasis on the evaluation of the net quark number response of the thermal medium to the presence of an external quark or antiquark sources, which we have recently advocated to provide a sensible probe of the net quark number content of the different regions of the phase diagram \cite{MariSurkau:2025MesBar}. We show how these quantities can disambiguate the interpretation of certain phases in cases where the behavior of the Polyakov loops is misleading, and how they allow one to scan the whole phase diagram while connecting various singular points, including the CEP, together. In Sec.~\ref{sec: thermodynamics}, in a more conventional analysis, we compute various thermodynamical observables of phenomenological interest, before concluding in Sec.~\ref{sec: conclusions}. Appendices B, C and D gather some more technical aspects.

\section{Framework}\label{sec: framework}
\subsection{The Nambu--Jona-Lasinio model}\label{sec: NJL}
The Nambu--Jona-Lasinio (NJL) model \cite{Nambu1961DynamicalI, Nambu1961DynamicalII} is one of the simplest models that exhibit spontaneous chiral symmetry breaking. Y. Nambu and G. Jona-Lasinio proposed it to study relativistic fermions under a strong interaction, and it is modeled after the BCS theory of superconductivity. It starts from the free quark part of the QCD Lagrangian density and adds a contact four-quark interaction term:
\begin{equation}\label{eq: general L}
    \L = \bar\psi\left(\slashed{\partial}+\hat{m}_0-\mu\gamma_0\right)\psi + \L_{\rm int}\,.
\end{equation}
This can be seen as simulating what QCD would look like upon integrating the gluon dynamics\footnote{We note that in the Curci-Ferrari model this is strongly motivated by the finite gluon mass, which is of the order of magnitude of the NJL cutoff, and the NJL model can be recovered as a low-energy limit of the model.}, leaving only a global U(1) charge symmetry related to quark/baryon number conservation, a global color symmetry associated with the quark internal degrees of freedom, and the chiral symmetry present in the massless QCD Lagrangian density. Here and in what follows, since our goal is to study the QCD phase diagram as a function of the temperature $T$ and quark chemical potential $\mu$, we use a formulation based on an Euclidean time restricted to the interval $[0,\beta]$, with $\smash{\beta\equiv 1/T}$ the inverse temperature. The chemical potential $\mu$ is the one associated with the conserved net quark number of the system. We also use Euclidean $\gamma_\nu$ matrices such that $\smash{\{\gamma_\mu,\gamma_\nu\}=2\delta_{\mu\nu}\mathds{1}}$, and define $\smash{\slashed{\partial}\equiv\gamma_\nu \partial_\nu}$. Finally, the quark fields $\psi$ have $N_f$ spinorial components $\smash{\psi\equiv(\psi_u,\psi_d,\psi_s,\dots)^{\rm t}}$ with $N_c$ color components each, while $\smash{\hat{m}_0\equiv\operatorname{diag\,}(m_u,m_d,m_s,\dots)}$ denotes the quark mass matrix.

The interaction term $\smash{\L_{\rm int}\sim(\bar\psi\Gamma\psi)^2}$ could \textit{a priori} include any of the terms that respect the global U(1) charge symmetry and the chiral symmetry SU$(N_f)_L\times$SU$(N_f)_R$ of the massless case.\footnote{It should also respect the remaining global color symmetry SU($N_c$) present without gluons, unless we allow for the formation of diquarks.} They are built from the (pseudo-) scalar, (pseudo-)vector, and tensor four-fermion contact interactions
\beq
    &&(\bar\psi\psi)^2, \,\ (\bar\psi\gamma_5\psi)^2, \,\ (\bar\psi\gamma_\mu\psi)^2, \,\ (\bar\psi\gamma_\mu\gamma_5\psi)^2, \,\ (\bar\psi i\sigma_{\mu\nu}\psi)^2,\nonumber\\
    &&(\bar\psi\vec{t}\psi)^2,\, (\bar\psi\gamma_5\vec{t}\psi)^2, \, (\bar\psi\gamma_\mu\vec{t}\psi)^2, \, (\bar\psi\gamma_\mu\gamma_5\vec{t}\psi)^2, \, (\bar\psi i\sigma_{\mu\nu}\vec{t}\psi)^2\!,\quad\nonumber
\eeq
where $\smash{\vec{t}=(t^1,t^2,\dots)}$ denotes the collection of SU($N_f$) generators and $\smash{\gamma_5\equiv\gamma_0\gamma_1\gamma_2\gamma_3}$. Depending on $N_f$, different combinations of these terms correspond to the different parts of the quark model. 

The simplest two-flavor model consists of a scalar and a pseudoscalar interaction term
\begin{equation}
    \L_{\rm int}=-\frac{G}{2}\left((\bar\psi\psi)^2+(\bar\psi i\gamma_5\vec{\tau}\psi)^2\right),
\end{equation}
made to describe chiral symmetry breaking through the quark condensate and the pseudoscalar mesons made from $u$ and $d$ quarks, the pions. Here, $\vec{\tau}$ denotes the collection of Pauli matrices, and isospin symmetry $\smash{m_u=m_d\equiv m_l}$ is usually assumed for the light quarks.\footnote{When looking at 2+1 flavors where the pseudoscalar nonet (pions, kaons, eta, eta') needs to be considered, one can use the same term while replacing the collection $\vec{\tau}$ of Pauli matrices by the collection $\vec{\lambda}$ of Gell-Mann matrices, but one should also add the six-quark t'Hooft anomaly term \cite{Kunihiro:1989my, Fu:2007xc, Fukushima:2008wg}.} 

For practical calculations, it is convenient to perform a Hubbard–Stratonovich transformation \cite{Stratonovich1957OnFunctions, Hubbard1959CalculationFunctions} to rewrite the Lagrangian density using auxiliary bosonic fields. More precisely, without changing the theory, one introduces auxiliary fields $\sigma$ and $\vec{\pi}$, corresponding to appropriately shifted, Gaussian fields. The total Lagrangian density, including these auxiliary fields, reads
\beq
{\cal L}_{\rm aux} & = & \bar\psi\left(\slashed{\partial}+m_l-\mu\gamma_0\right)\psi-\frac{G}{2}\Big[(\bar\psi\psi)^2+(\bar\psi i\gamma_5\vec{\tau}\psi)^2\Big]\nonumber\\
& & +\,\frac{1}{2G}(\sigma+G\bar\psi\psi)^2+\frac{1}{2G}(\vec{\pi}+G\bar\psi i\gamma_5\vec{\tau}\psi)^2\,.
\eeq
The shifts of $\sigma$ and $\vec{\pi}$ are tailor-made such that the four-quark interaction terms cancel, and we are left with
\beq
{\cal L}_{\rm aux} & = & \bar\psi\left(\slashed{\partial}+m_l-\mu\gamma_0\right)\psi+\frac{\sigma^2+\vec{\pi}^2}{2G}\nonumber\\
& & +\,\sigma\bar\psi\psi+\vec{\pi}\cdot \bar\psi i\gamma_5\vec{\tau}\psi\,.\label{eq:sigma}
\eeq
We stress that there are no kinetic terms for the $\sigma$ and $\vec{\pi}$ fields.

For two degenerate flavors, the Lagrangian density is invariant under SU$(2)_V$ transformations corresponding to
\beq
\psi\to e^{i\vec{\theta}\cdot\vec{\tau}}\psi,\,\,\,\bar\psi\to\bar\psi e^{-i\vec{\theta}\cdot\vec{\tau}}\!,\,\,\,\sigma\to\sigma,\,\,\,\vec{\pi}\to{\cal R}_{\vec{\theta}}\,\vec{\pi},
\eeq
where ${\cal R}_{\vec{\theta}}$ denotes the rotation of angle $\smash{\theta\equiv||\vec{\theta}||}$ and axis $\smash{\vec{n}\equiv\vec{\theta}/\theta}$. This symmetry is not broken spontaneously, which implies in particular that the expectation value $\langle\vec{\pi}\rangle$ vanishes. In the chiral limit, that is when $\smash{m_l\to 0}$, the Lagrangian density is also invariant under SU$(2)_A$ transformations corresponding to
\beq
\psi\to e^{i\gamma_5\vec{\theta}\cdot\vec{\tau}}\psi,\,\,\, \bar\psi\to\bar\psi e^{i\gamma_5\vec{\theta}\cdot\vec{\tau}},\,\,\,
\sigma\to-\sigma,\,\,\,\vec{\pi}\to\vec{\pi}.
\eeq
{\it A priori,} this would imply that the expectation value $\langle\sigma\rangle$ vanishes in this case. However, the SU$(2)_A$ symmetry is known to be broken spontaneously at low temperatures, leading to a non-vanishing expectation value $\langle\sigma\rangle$ which plays the role of a dynamically generated mass for the quarks. As the temperature is increased, there is a temperature above which $\langle\sigma\rangle$ becomes equal to $0$, signaling the restoration of the symmetry and making $\langle\sigma\rangle$ an order parameter for chiral symmetry, the so-called chiral condensate. In the presence of a small but non-zero quark mass $m_l$, the symmetry is explicitly broken, but the expectation value $\langle\sigma\rangle$ remains dominant over $m_l$ at low temperatures and crossovers to a negligible value at higher temperatures. We shall then still refer to $\langle\sigma\rangle$ as an order parameter in what follows.

If one is interested only in $\langle\sigma\rangle$ as it drives the chiral symmetry breaking transition, it is possible to integrate the quark fields exactly. This leads to the effective Lagrangian density
\beq\label{eq:Leff}
{\cal L}_{\rm eff}=\frac{\sigma^2+\vec{\pi}^2}{2G} -\ln\det\Big(\slashed{\partial}+m_l+\sigma-\mu\gamma_0+i\gamma_5\vec{\tau}\cdot\vec{\pi}\Big)\,.\nonumber\\
\eeq
In principle, to study the order parameter behavior of $\langle\sigma\rangle$, one should introduce sources coupled linearly to $\psi$, $\bar\psi$, $\sigma$, $\vec{\pi}$, Legendre transform the logarithm of the corresponding partition function, and look for extrema of the resulting effective action, corresponding to the limit of zero sources. However, since we know that $\langle\psi\rangle$, $\langle\bar\psi\rangle$, and $\langle\vec{\pi}\rangle$ all vanish in this limit, one can more conveniently consider a reduced effective action where quark, antiquark, and pion fields have been set to $0$. Moreover, assuming translation invariance, we can restrict to homogeneous configurations of $\sigma$ and compute the corresponding effective potential $V_{\rm NJL}(\sigma)$. 

At tree-level of the theory (\ref{eq:Leff}), the effective potential is just given by the classical action, and we then find the well-known result\footnote{More precisely, one obtains $-4N_fN_c\int_Q^T \ln {\rm det}\,(-i\hat Q_\mu\gamma_\mu+M)$. However, using that $\gamma_5^2=\mathds{1}$ and $\gamma_5\gamma_\mu\gamma_5=-\gamma_\mu$, this rewrites $-4N_fN_c\int_Q^T \ln {\rm det}(i\hat Q_\mu\gamma_\mu+M)$. Averaging over these two equivalent expressions, one arrives at Eq.~\eqref{eq:VNJL}.}
\beq
V_{\rm NJL}(\sigma)=\frac{\sigma^2}{2G}-2N_fN_c\int_Q^T \ln\,(\hat Q^2+M^2)\,,\label{eq:VNJL}
\eeq
with the constituent quark mass\footnote{Note that we use $\langle\bar\psi\psi\rangle=2\langle\bar\psi_l\psi_l\rangle$, the sum of the two (degenerate) light quark condensates.}
\begin{equation}\label{eq: const M}
    M\equiv m_l+\sigma= m_l-G\langle\bar\psi\psi\rangle\,.
\end{equation}
Our notations are such that $\smash{\hat Q\equiv(\hat\omega_n-i\mu,\vec{q})}$, with $\smash{\hat\omega_n\equiv2\pi(n+1/2)T}$ a fermionic Matsubara frequency and
\beq
\int_{\hat Q}^T f(\hat Q)\equiv T\sum_{n\in\mathds{Z}}\int\frac{d^3q}{(2\pi)^3}f(\hat\omega_n-i\mu,\vec{q})\label{eq:sum0}
\eeq
the corresponding Matsubara sum-integral, where it is implicitly understood that the $q$-integral is cut off at a scale $\Lambda$, which is part of the parameters of the non-renormalizable NJL model, together with $m_l$ and $G$.

To fix the parameters, one needs to reproduce physical quantities within the model. For the two-flavor case, two obvious choices are the vacuum pion mass $m_\pi$ and vacuum pion decay constant $f_\pi$. The third parameter may be fixed by imposing a value on the vacuum quark condensate. Determining $G,\,\Lambda$ and $m_l$ from vacuum physics makes the finite temperature results proper predictions of the model. The current mass $m_l$ can directly be determined through the Gell-Mann, Oakes, Renner (GMOR) relation \cite{Gell-Mann1968GMOR}
\begin{equation}\label{eq: GMOR}
    m_\pi^2 f_\pi^2=-m_l\langle\bar\psi\psi\rangle,
\end{equation}
which holds near the chiral limit. The trace over the quark propagator gives the vacuum quark condensate
\begin{equation}\label{eq: vac condensate}
    \langle\bar\psi\psi\rangle 
    =-8N_c\int\frac{d^4p}{(2\pi)^4}\frac{M}{p^2+M^2}\theta(\Lambda^2-\vec{p}^2)\,,
\end{equation}
corresponding to the gap equation $\smash{\partial V_{\rm NJL}/\partial\sigma=0}$ in the vacuum ($\smash{T=\mu=0}$) with $V_{\rm NJL}$ given in Eq.~\eqref{eq: const M}. The pion decay constant is given by the vacuum to pion axial-vector matrix element
\begin{equation}\label{eq: pi decay}
    f_\pi^2 = 4N_c\int\frac{d^4p}{(2\pi)^4}\frac{M^2}{(p^2+M^2)^2}\theta(\Lambda^2-\vec{p}^2)\,.
\end{equation}
The parameters $G$ and $\Lambda$ are then fixed by simultaneous resolution of Eqs.~\eqref{eq: vac condensate} and \eqref{eq: pi decay}, after plugging in Eqs.~\eqref{eq: const M} and \eqref{eq: GMOR} and imposing values for the vacuum pion physics. In two-flavor applications related to the phase diagram, it has been customary \cite{Fukushima:2003fw, Ratti:2005jh, Roessner:2006xn} to just fix a current quark mass $\smash{m_l=5.5\,}$MeV and leave the condensate free, as comparatively large errors are associated with it. With that, we employ a common parameter choice \cite{Hatsuda:1994pi, Klevansky:1992qe, Meisinger:1995ih, Fukushima:2003fw} $\smash{\Lambda=631\,}$MeV, and $\smash{G=10.99\,\text{GeV}^{-2}}$, corresponding to $\smash{m_\pi\simeq138.2\,}$MeV, $\smash{f_\pi\simeq93.1\,}$MeV and  $\smash{\langle\bar\psi_l\psi_l\rangle^{1/3}\simeq -247\,}$MeV. Moreover, with these parameter values, the vacuum quark mass is found to be $\smash{M_0= M(T=0,\mu=0)\simeq335.9\,}$MeV. We note that the quark condensate is more constrained nowadays, and that, in the isospin-symmetric case and in the absence of QED corrections, a pion mass of ${m_\pi\simeq135\,}$MeV may be better motivated and is usually used in Lattice QCD calculations. We analyze the impact of different parameter choices in App.~\ref{app: params}, but the main text uses these parameters, for clearer comparison with previous two-flavor PNJL studies \cite{Fukushima:2003fw, Ratti:2005jh, Roessner:2006xn} and to highlight the effects of our modeling of the Polyakov loop potential.

\subsection{PNJL model}\label{sec: PNJL}
As is well known, another major actor of QCD at finite temperature is the confinement-deconfinement transition, as probed by the other order parameter of QCD, the Polyakov loop. In some appropriate gauges, it is well understood that a non-trivial background or gauge-field expectation value $\langle A_\mu^a\rangle$ drives the behavior of the Polyakov loop. This means that, if one wants to incorporate the physics of the confinement/deconfinement transition into the NJL model, one needs to separate the relevant background from the gauge field before integrating over the gluonic fluctuations. In practice, this means replacing the normal derivative $\slashed{\partial}$ in Eq.~\eqref{eq:sigma} by a covariant derivative $\smash{\slashed{D}=\slashed{\partial}-ig\langle A_\mu^a\rangle t^a}$ in the presence of the background $\langle A_\mu^a\rangle$.

In specific gauges, the gauge-field expectation value takes a simple enough form allowing one to easily assess the impact on the potential \eqref{eq:VNJL}. In particular, in cases where the gauge-field expectation value is constant, temporal and along the diagonal directions of the SU(3) algebra
\beq
\langle g A_\mu^a(x)\rangle=T\left(r_3\frac{\lambda_3}{2}+r_8\frac{\lambda_8}{2}\right),\label{eq:vevA}
\eeq
where the $\lambda_j$ are the Gell-Mann matrices, the potential (\ref{eq:VNJL}) generalizes to
\beq
V_{\rm PNJL}(\sigma,r_3,r_8)=\frac{\sigma^2}{2G}-2N_f\sum_\rho\int_Q^T \ln(\hat Q^2_\rho+M^2)\,,\label{eq: VPNJL}\nonumber\\
\eeq
with $\smash{\hat Q_\rho=(\hat \omega_n+T(r\cdot\rho)-i\mu)}$. In this formula, the labels $\rho$ denote the defining weights of SU$(N_c)$, corresponding to a sophisticated but very useful way to label the colors in the defining representation. For $\smash{N_c=3}$, these are the two-dimensional, real-valued vectors $1/2(1,1/\sqrt{3})$, $1/2(-1,1/\sqrt{3})$ and $(0,-1/\sqrt{3})$. The variable $r$ is also a vector $\smash{r\equiv(r_3,r_8)}$ and we have introduced the short-hand notation $\smash{r\cdot\rho\equiv r_3\rho_3+r_8\rho_8}$. 

What the formula \eqref{eq: VPNJL} tells in essence is that the presence of a non-trivial background of the form \eqref{eq:vevA} lifts the degeneracy of the various color modes by introducing a color-dependent shift of the frequency, or, better, a color-dependent imaginary shift of the chemical potential. In the absence of a background, the various color modes become degenerate, and one recovers Eq.~\eqref{eq:VNJL} with the factor of $N_c$ in front of the sum-integral.

\subsection{Center-symmetric Curci-Ferrari model}

To complete the model, we still need to specify in which gauge the gauge-field expectation values $r_3$ and $r_8$ are computed. Here, we choose the center-symmetric Landau gauge (csLg) introduced in Ref.~\cite{vanEgmond:2021jyx}. The benefit of this gauge is that the corresponding effective potential $V_{\rm csLg}(r_3,r_8)$, the extremization of which gives the expectation values $r_3$ and $r_8$ in the pure glue case, is invariant under particular transformations of $\smash{r\equiv (r_3,r_8)}$, corresponding to the center transformations that govern the confinement/deconfinement transition in the pure gauge sector.

A major problem of this gauge, however, as most Lorentz covariant gauges, is that the Faddeev-Popov gauge-fixing procedure is ambiguous at low energies. In this regime, one needs to model the gauge-fixing procedure based on lattice simulations of the same gauge. Such simulations, which are able to elude the ambiguity, exist for the center-symmetric Landau gauge \cite{vanEgmond:2024ljf, vanEgmond:2025zxf} and show in particular that the gluon propagator is screened at low energies. A minimal modeling of this fact consists of incorporating a mass term to the partially gauge-fixed Faddeev-Popov action. To preserve center-symmetry, the mass term is introduced in a specific way. This leads to the center-symmetric CF model \cite{vanEgmond:2021jyx}.

Therefore, in what follows, rather than computing $V_{\rm csLg}(r_3,r_8)$, which is not accessible to date due to the gauge-fixing ambiguity, we consider the corresponding CF potential $V_{\rm csCF}(r_3,r_8)$, which has been shown to provide a good grasp on certain aspects of the confinement/deconfinement transition. Moreover, another interesting feature of the CF model is that the screening of the gluon tames the infrared singularities, making perturbation theory viable at low energies in the gluonic sector \cite{Tissier:2011ey, Reinosa:2014ooa,Pelaez:2021tpq}. We will then consider $V_{\rm csCF}(r_3,r_8)$ at one-loop order as originally computed in Ref.~\cite{vanEgmond:2021jyx}, see also Ref.~\cite{MariSurkau:2024zfb} for more details. Formally, it writes
\beq
V_{\rm csCF}(r_3,r_8) & = & \frac{m^2}{2g^2}(r-\bar r)^2\nonumber\\
& + & \frac{d-2}{2}\sum_\kappa\int_Q^T\ln\big[Q_\kappa^2+m^2\big]\nonumber\\
& + & \frac{1}{2}\sum_\kappa\int_Q^T\ln\left[1+\frac{m^2\bar Q_\kappa^2}{(Q_\kappa\cdot\bar Q_\kappa)^2}\right].\label{eq: VcsLg}
\eeq
Our notations are such that $\smash{Q_\kappa=(\omega_n+T(r\cdot\kappa),\vec{q})}$ and $\smash{\bar Q_\kappa=(\omega_n+T(\bar r\cdot\kappa),\vec{q})}$ with $\smash{\omega_n\equiv 2\pi nT}$ a bosonic Matsubara frequency and
\beq
\int_Q^T f(Q)\equiv T\sum_{n\in\mathds{Z}}\int\frac{d^{d-1}q}{(2\pi)^{d-1}}f(\omega_n,\vec{q})\label{eq:sum}
\eeq
the corresponding Matsubara sum-integral. The labels $\kappa$ stand for the adjoint weights of SU$(N_c)$, similar to the defining weights that enter Eq.~\eqref{eq: VPNJL}. The difference is that there are $N_c-1$ vanishing such weights. As for the non-vanishing ones, known as roots, they can be obtained as all possible differences of distinct defining weights. In SU(3), they correspond to the vectors $\pm(1,0)$, $\pm(1/2,\sqrt{3}/2)$ and $\pm(1/2,-\sqrt{3}/2)$. Finally, $\bar r$ takes the specific value $(4\pi/3,0)$. This ensures that $V_{\rm csCF}(r_3,r_8)$ is center-symmetric and thus that its extremum plays the role of an order parameter for the confinement deconfinement/transition \cite{vanEgmond:2021jyx}.

Note that, in Eq.~\eqref{eq:sum}, we use dimensional regularization rather than a cut-off as we did in Eq.~\eqref{eq:sum0}. This is to make full use of the symmetries of the center-symmetric CF model. This model is renormalizable, unlike the NJL model. Of course, one could criticize the fact of coupling a renormalizable model to a non-renormalizable one. This, in a sense, should be taken as part of our modeling. However, it is still true that the renormalization procedure introduces an arbitrariness in how the two models are coupled, since one could compute $V_{\rm csCF}(r_3,r_8)$ in different schemes or at various renormalization scales. Sensible results should not be very sensitive to these choices.

For the reader's convenience, we give the final formula for the SU(3) potential in the center-symmetric gauge found by evaluating the sums/integrals in \eqref{eq: VcsLg} with ${\bar r=(4\pi/3,0)}$, for the general case see \cite{MariSurkau:2024zfb}, which can directly be implemented numerically
\begin{widetext}
    \beq
        V_{\rm csCF}(r_3,r_8)
        & = & -\frac{4\pi^2T^4}{405}+\frac{3T}{2\pi^2}\int_0^\infty \!\!dq\,q^2\,\ln\Big[\left(1-e^{-\varepsilon_q/T}\right)^2\left(1+e^{-\varepsilon_q/T}+e^{-2\varepsilon_q/T}\right)^3\Big]\nonumber\\
        & + & \frac{T^2}{2g^2}m^2\Delta r^2(1+\delta_{\rm scheme}) + \frac{T^4}{48}\Delta r^2\left(1+\frac{52\pi^2}{45}\frac{T^2}{m^2}\right) + \frac{T^4}{96\pi^2}\left(9(\Delta r^2)^2+8\pi(r_3-4\pi/3)(\Delta r^2-4{r_8}^2)\right) \nonumber\\
        & + & \frac{3T^2}{8\pi^2}\Delta r^2\int_0^\infty dq\,\frac{q^2}{\varepsilon_q}\left(3\frac{m^2}{q^2}+6+\frac{q^2}{m^2}\right)\frac{2+e^{\varepsilon_q/T}}{1+e^{\varepsilon_q/T}+e^{2\varepsilon_q/T}}\nonumber\\
        & + & \frac{T}{\pi}\sum_\kappa\sum_{n\in\mathds{Z}}\bigg[\frac{1}{12}\Big(2X_0^3 + 3\bar X^3 - \bar X_0^3 - 2X^3 - X_+^3 - X_-^3 \Big)-\,\frac{T}{4}(\kappa\cdot\Delta r)\,\bar\omega_n^\kappa\big( \bar X_0-3\bar X \big)\nonumber\\
        & & \hspace{1.0cm}   +\,\frac{T^2}{4}(\kappa\cdot\Delta r)^2\bar X+\frac{T^3}{6}(\kappa\cdot\Delta r)^3{\rm sgn}(\bar\omega^\kappa_n) +\,\frac{T^2}{16}(\kappa\cdot\Delta r)^2(\bar\omega_n^\kappa)^2\left(\frac{6}{\bar X}-\frac{1}{\bar X_0}-\frac{2}{\bar X_0 + \bar X}\right)\bigg]\,.\label{eq: VcsCF}
    \eeq
\end{widetext}
The quantities appearing in the last sum are
\beq
    X_0 &\equiv& \sqrt{\omega_n^\kappa\bar\omega_n^\kappa}, \phantom{|\bar\omega_n^\kappa|} X \equiv \sqrt{(\omega_n^\kappa)^2+m^2},\\
    \bar X_0 &\equiv& |\bar\omega_n^\kappa|, \phantom{\sqrt{\omega_n^\kappa\bar\omega_n^\kappa}}  \bar X \equiv \sqrt{(\bar{\omega}_n^\kappa)^2+m^2}, \\
    X_\pm &\equiv& \sqrt{X_0^2+\frac{m^2}{2}\Bigg(1\pm\sqrt{1+4\frac{X_0^2-\bar X_0^2}{m^2}}\Bigg)}\,,
\eeq
and we have introduced the shifted, bosonic frequencies $\smash{\omega_n^\kappa\equiv (2\pi n+\kappa\cdot r)T}$ and ${\bar\omega_n^\kappa\equiv (2\pi n+\kappa\cdot \bar r)T}$, the gluon quasiparticle energy ${\varepsilon_q=\sqrt{q^2+m^2}}$, and the difference $\smash{\Delta r=r-\bar r=(r_3-4\pi/3,r_8)}$ with the square being $\smash{{\Delta r}^2=(r_3-4\pi/3)^2+{r_8}^2}$. The first line corresponds to a purely background contribution and disappears when extremizing the potential, but contributes to the correct Stefan-Boltzmann limit at high $T$. 

In what follows, we choose to work within one popular scheme known as infrared-safe (IRs), where perturbation theory is defined at all scales \cite{Tissier:2011ey}, corresponding to 
\begin{equation}
    \delta_{\rm IRs}=-\frac{3g^2}{64\pi^2t^2}\Big(t+\frac{5}{2}t^2+t^3\ln(t)-(1+t)^2\ln(1+t)\Big),
\end{equation}
where ${t=\xi^2/m^2}$, with the renormalization scale $\xi$. We present our results at a renormalization scale of $\smash{\xi=1\,\text{GeV}\approx2\pi 160\,}$MeV, which is roughly the first (gluonic) Matsubara frequency $\smash{\omega_1=2\pi T}$ near the transition at vanishing chemical potential, and thus considered a natural scale. At this scale, the CF gluon mass parameter is $\smash{m\simeq0.39\,}$GeV, and the critical deconfinement transition temperature found with the one-loop potential \eqref{eq: VcsLg} in the quenched (Yang-Mills) theory is $\smash{T_{\rm c}^{\rm YM}\simeq253\,}$MeV \cite{MariSurkau:2024zfb}, quite close to the lattice expectation of $\smash{\sim270\,}$MeV for a one-loop approximation. In App. \ref{app: params}, we study the impact of the renormalization scale choice on our results, and find that it barely matters, confirming the consistency of the approach.

We study the interplay between the chiral and confinement/deconfinement transitions using the total potential obtained by combining \eqref{eq: VPNJL} and \eqref{eq: VcsLg}/\eqref{eq: VcsCF}
\beq
V(\sigma,r_3,r_8)\equiv V_{\rm PNJL}(\sigma,r_3,r_8)+V_{\rm csCF}(r_3,r_8)\,.\label{eq:model}
\eeq
The phase diagram of the model is obtained by studying, as functions of $T$ and $\mu$, the order parameters $\sigma$ and $(r_3,r_8)$ obtained at the extremum of the potential
\begin{equation}\label{eq: saddle pt}
    0=\frac{\partial V}{\partial\sigma}=\frac{\partial V}{\partial r_3}=\frac{\partial V}{\partial r_8}.
\end{equation}
From the knowledge of $(r_3,r_8)$, one deduces the gauge-invariant order parameters $\ell$ and $\bar\ell$. At leading order, the relation between $(r_3,r_8)$ and $(\ell,\bar\ell)$ is found to be
\beq
\ell & = & \frac{e^{-i\frac{r_8}{\sqrt{3}}}+2e^{i\frac{r_8}{2\sqrt{3}}}\cos(r_3/2)}{3}\,,\label{eq:l0}\\
\bar\ell & = & \frac{e^{i\frac{r_8}{\sqrt{3}}}+2e^{-i\frac{r_8}{2\sqrt{3}}}\cos(r_3/2)}{3}\,.\label{eq:lb0}
\eeq
We mention, however, that, because \eqref{eq: VcsLg} corresponds in fact to a next-to-leading order calculation, it is more consistent to use the next-to-leading order relation between $(r_3,r_8)$ and $(\ell,\bar\ell)$ which we derive in App.~\ref{app: NLO_l}. We will show both results for the Polyakov loops below. 

Finally, we note that, for applications involving a non-zero, real $\mu$, while $r_3$ is real, $r_8$ is purely imaginary and non-zero, corresponding to both $\ell$ and $\bar\ell$ real and distinct \cite{MariSurkau:2025dfo}.

\begin{figure}[t]
    \centering
    \includegraphics[width=0.9\linewidth]{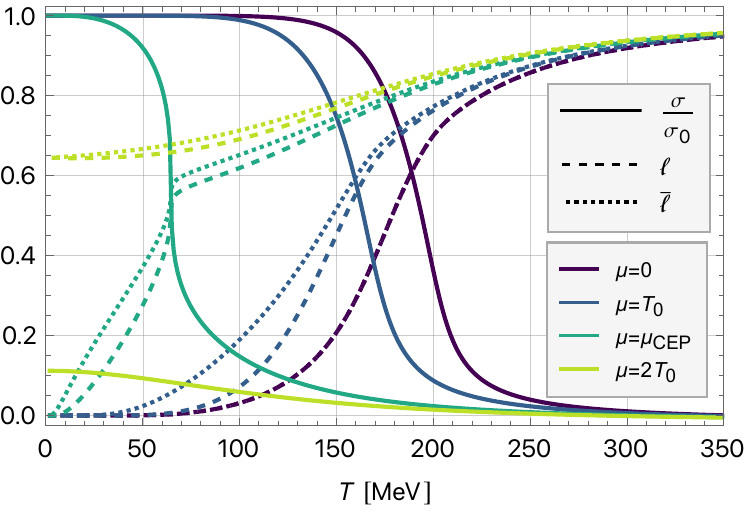}
    \caption{Order parameters as functions of the temperature $T$ at various chemical potentials $\mu$. Here $T_0\simeq186\,$MeV stands for $(T_{\rm pc}^\sigma+T_{\rm pc}^\ell)/2$ at $\smash{\mu=0}$, see the main text for the definitions of $T_{\rm pc}^\sigma$ and $T_{\rm pc}^\ell$, and $\mu_{\rm CEP}\simeq324\,$MeV.}
    \label{fig: ord par}
\end{figure}

\section{Phase diagram\label{sec: phase dia}}

The different regions of the phase diagram are delimited by studying the behavior and changes thereof of the order parameters at finite temperatures and chemical potentials. In Fig.~\ref{fig: ord par}, we show the behavior of $\sigma$, $\ell$, and $\bar\ell$ as functions of the temperature and for different values of the quark chemical potential. Since there is no exact symmetry protecting these quantities, there is no range of temperatures over which they acquire a fixed value characteristic of true order parameters. However, over a wide range of values of $\mu$, including $\smash{\mu=0}$, these quantities behave as approximate order parameters in the sense that they transition between two extremal, and rather different values as $\smash{T\to 0}$ and $\smash{T\to\infty}$.

In this crossover region, one can define a pseudo-critical transition temperature $T_{\rm pc}(\mu)$ from the inflection of the corresponding order parameters
\begin{equation}
\left.\frac{\partial^2\sigma}{\partial T^2}\right|_{T_{\rm pc}^\sigma}=0, \quad    \left.\frac{\partial^2\ell}{\partial T^2}\right|_{T_{\rm pc}^\ell}=0, \quad \mbox{or} \quad \left.\frac{\partial^2\bar\ell}{\partial T^2}\right|_{T_{\rm pc}^{\bar\ell}}=0,\label{eq:Tpc}
\end{equation}
i.e., the peak of their temperature susceptibilities. \textit{A priori}, the pseudo-critical temperatures associated with different order parameters can differ. In the present case, however, we find that the crossovers all lie rather close to each other and that they happen in a fairly narrow temperature window (narrower for the chiral condensate than for the Polyakov loops), as visible in Fig.~\ref{fig: ord par}.

\subsection{Vanishing chemical potential}
At vanishing chemical potential, the pseudo-critical temperatures $T_{\rm pc}^\ell$ and $T_{\rm pc}^{\bar\ell}$ coincide. This is expected since, owing to charge conjugation invariance, $r_8$ should vanish at the extremum of the potential \cite{MariSurkau:2025dfo}, and, thus, $\ell$ and $\bar\ell$ should coincide, which is indeed what we find, see the darkest dashed line in Fig.~\ref{fig: ord par}. This common temperature $\smash{T_{\rm pc}^\ell\simeq175}$\,MeV is only about $20$\,MeV away from the chiral pseudo-critical temperature $\smash{T_{\rm pc}^\sigma\simeq197}$\,MeV. These values agree roughly with those (of the order of $\smash{200\text{-}220\,}$MeV) obtained from PNJL models based on other glue potentials \cite{Fukushima:2003fw, Ratti:2005jh, Roessner:2006xn}, but we note that we naturally find slightly lower values, which is desirable given the values of around $175\,$MeV found in lattice simulations with two flavors \cite{Bornyakov:2009qh, CP-PACS:2000phc, Karsch:2000kv, Karsch:2001vs}. Note also that we did not employ the popular procedure of lowering the scale of the gauge potential by hand to achieve a better agreement with lattice results \cite{Ratti:2005jh, Roessner:2006xn, Schaefer:2007pw, Haas:2013qwp, Herbst:2013ail}. For a more detailed analysis of the impact of varying the parameters, see App.~\ref{app: params}, where we show that much closer agreement with lattice results is in fact achievable using well-motivated parameter changes.

Another interesting fact is that, had we used the self-consistent background potential of Ref.~\cite{Reinosa:2014ooa} instead of the center-symmetric Landau gauge potential (\ref{eq: VcsLg}), the difference between the chiral and deconfinement pseudo-critical temperatures would have been much larger, of the order of 80\,MeV. In this case, the deconfinement crossover happens already at an unrealistically low temperature of 101 MeV, and the Polyakov loops saturate at $\smash{\ell=1}$ at around ${T\simeq142\,}$MeV, which is an artifact of the one-loop treatment in the self-consistent background gauge. The PNJL model thus returns to the usual NJL model at this temperature, and the chiral crossover for the chosen parameter values happens around 178 MeV. The improvement observed here upon using the center-symmetric Landau gauge potential is consistent with similar improvements observed in other finite-temperature applications \cite{MariSurkau:2024zfb, MariSurkau:2025dfo}.

 \begin{figure}[t]
    \centering
    \includegraphics[width=0.9\linewidth]{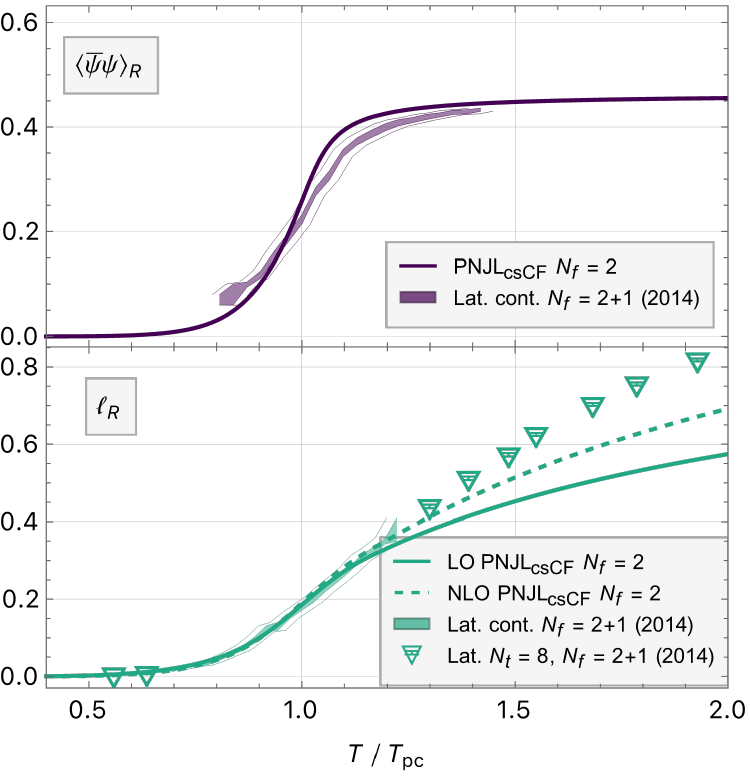}
    \caption{Comparison of the temperature dependence of the renormalized chiral condensate $\langle\bar\psi\psi\rangle_R$ (top) and LO and NLO Polyakov loop $\ell_R$ (bottom) found in this work (lines, ${N_f=2}$) to continuum extrapolated lattice data (bands), and data from a single lattice size with temporal extent $N_t=8$ (triangles), both from Ref.~\cite{Borsanyi:2010ordPar} (${N_f=2+1}$). The temperature has been rescaled by the pseudocritical temperature for each observable, see the discussion in the main text. The thin lines around the lattice data bands (horizontal error) show the 2\% systematic temperature uncertainty due to the scale setting.}
    \label{fig: ord par vs lat dat}
\end{figure}

In Figure \ref{fig: ord par vs lat dat}, we show our best attempt at comparing the order parameters obtained in this work at $\smash{\mu=0}$ with results of lattice simulations. To the best of our knowledge, there exists no lattice data for ${N_f=2}$ QCD with realistic pion masses, so we resort to a comparison with the continuum extrapolated ${N_f=2+1}$ data from Ref.~\cite{Borsanyi:2010ordPar}.\footnote{The authors of Ref.~\cite{Roessner:2006xn} compared their results for the Polyakov loop to ${N_f=2}$ heavy-quark lattice QCD results \cite{Kaczmarek:2005ui} with an unrealistically large pion mass, ${m_\pi\sim\mathcal{O}(770\,}$MeV) on a single ${16^3\times4}$ lattice, where the crossover of $\ell_R$ is steeper. Our results for the Polyakov loop also compare fine with this data.} Since we are comparing theories with different quark contents, the order parameters should not behave exactly the same, but we expect that the strange quark contribution is less significant than that of the two light quarks, in particular for temperatures below the strange quark mass ${m_s\simeq 100\,}$ MeV. The authors of Ref.~\cite{Borsanyi:2010ordPar} define the renormalized chiral condensate as
\begin{equation}\label{eq: renorm chir cond}
    \langle\bar\psi\psi\rangle_R=\frac{m_l}{m_\pi^4}\left(\langle\bar\psi\psi\rangle_{T=0}-\langle\bar\psi\psi\rangle_T\right),
\end{equation}
which we can directly compute from $\sigma(T)$ using the parameters $m_l$, $G$ and the pion mass $\smash{m_\pi\simeq138.2}\,$MeV found with them, see Sec.~\ref{sec: NJL}. Since the crossover temperature $\smash{T_{\rm pc}^\sigma\simeq197\,}$MeV found in the present model differs by ${\sim25\%}$ from the one in the lattice simulations, $\smash{T_{\rm pc,\,lat}^{\langle\bar\psi\psi\rangle_R}\simeq155\,}$MeV, we perform the comparison of our results and the data in terms of $T/T_{\rm pc}$, with $T_{\rm pc}$ the corresponding chiral crossover temperature in each case. We find that the temperature evolution through the crossover matches fairly well, but the present results have a slightly steeper behavior after the crossover transition. We also note that the situation improves significantly if, instead of fixing the NJL parameters using $\smash{m_\pi=138}\,$MeV, we choose $\smash{m_\pi=135}\,$ MeV instead, see App.~\ref{app: params} for details. The limiting value at high temperatures is also well reproduced, where we stress that no vertical rescaling (i.e. of $\langle\bar\psi\psi\rangle_R$) was done since we use the same definition \eqref{eq: renorm chir cond}.

As for the Polyakov loop, we need to allow for a rescaling corresponding to a change of renormalization scheme, since the Polyakov loop renormalizes and Polyakov loops computed within different schemes coincide only after an appropriate, temperature-dependent rescaling ${\ell_R=e^{-\beta\alpha}\ell}$, with a constant $\alpha$ and ${\beta=1/T}$. This form arises from the freedom in choosing the zero point of the free energy change induced by the static quark source. In Fig.~\ref{fig: ord par vs lat dat} we use ${\alpha^{(0)}\simeq0.181\,}$GeV and ${\alpha^{(1)}\simeq0.385\,}$GeV for the leading-order and next-to-leading-order Polyakov loops respectively, see below. Note that no Polyakov loop crossover temperature is given for the continuum extrapolated lattice results \cite{Borsanyi:2010ordPar}, as the crossover is too wide and the results are not extrapolated to temperatures above 220\,MeV. For the sake of showing the Polyakov loop results in the same figure as those for the chiral condensate, we rescale the lattice data with the temperature ${T_{\rm pc,\, lat.}^{\ell_R}\simeq180\,}$MeV, which seems a reasonable value considering the behavior of the data available, and ours by ${T_{\rm pc,\, PNJL}^{\ell_R^{(0)}}\simeq179.5\,}$MeV and ${T_{\rm pc,\, PNJL}^{\ell_R^{(1)}}\simeq182.5\,}$MeV.

Our results agree well with the data up to $\smash{T/T_{\rm pc}\simeq 1.2}$. It should be noted that this is only the case after rescaling to match the lattice renormalization scheme. Before, the crossover found with the PNJL model is significantly stronger (cf. Fig.~\ref{fig: ord par}) than that on the lattice, which is the case rather independently of the Polyakov loop potential used. However, even after fitting the continuum extrapolated lattice data, a discrepancy persists at temperatures above ${T\gtrsim1.2T_{\rm pc}}$ compared to the ${N_t=8}$ lattice data. Various reasons could explain this mismatch, from the missing strange quark in our approach to the fact that the lattice results we are comparing to are not extrapolated to the continuum beyond $T_{\rm pc}$. Also, in most PNJL approaches, non-Gaussian thermal fluctuations that lead to a temperature-dependent renormalization of the Polyakov loop on the lattice are not included, although they are known to play a role \cite{Herbst:2015ona}. Actually, the lattice computes the Polyakov loops as the average ${\langle\Phi[A]\rangle}$ of some functional $\Phi[A]$, while most models compute ${\Phi(\langle A\rangle)}$  instead. In the present approach, we can compute corrections beyond this semi-classical approximation. One-loop corrections (referred to as NLO in Fig.~\ref{fig: ord par vs lat dat}) are evaluated in App.~\ref{app: NLO_l}, and although they cannot entirely account for the mismatch with the lattice data, we see that they tend to reduce the discrepancy. At higher temperatures further thermal fluctuations may be more efficiently included via a Hard Thermal Loop (HTL) resummation.

\subsection{Finite chemical potential and CEP}
\begin{figure}[t]
    \centering
    \includegraphics[width=0.86\linewidth]{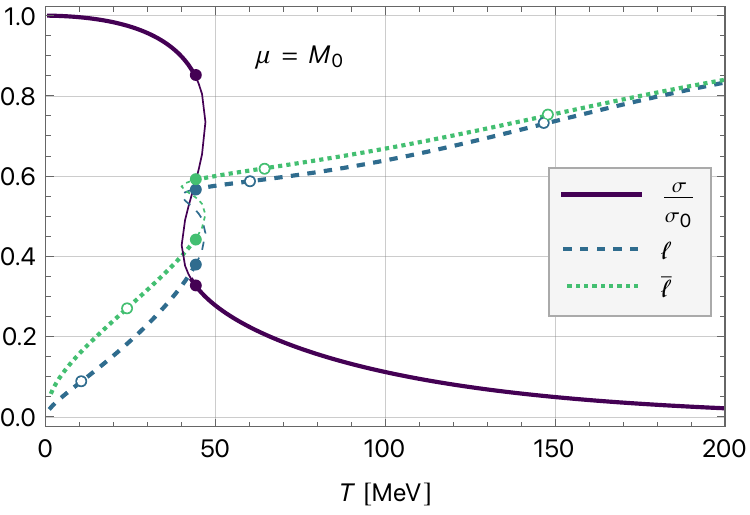}
    \caption{Temperature dependence of the order parameters at the chemical potential ${\mu=M_0\simeq336\,}$MeV corresponding to the vacuum constituent quark mass, which crosses through the spinodal region. Thin lines represent metastable and unstable extrema. The filled points mark the first-order transition, and the open points mark additional inflection points, see the text and App.~\ref{app:inflections} for more details.}
    \label{fig: ord pars at mu=M}
\end{figure}

As the chemical potential is increased away from $\smash{\mu=0}$, $T_{\rm pc}^\ell$ and $T_{\rm pc}^{\bar\ell}$ become different, signaling the explicit breaking of charge conjugation, but they remain pretty close to each other, with their gap never exceeding 2.5\,MeV. The gap between these temperatures and the chiral pseudo-critical temperature shrinks until one reaches a critical endpoint located at $\smash{(\mu_{\rm CEP}, T_{\rm CEP})\simeq(324,64)\,}$MeV, after which the transition becomes first order, with a discontinuous jump of all three order parameters at the same temperature $T_c(\mu)$. The first order transition curve $\mu_c(T)$ extends from $\smash{\mu_{\rm CEP}=\mu_c(T_{\rm CEP})}$ to $\smash{\mu_c(0)\equiv\mu_c(T=0)\simeq 348\,}$MeV. It is surrounded by a spinodal region, originating at the CEP and widening towards the $\mu$-axis, where it is almost 11\,MeV wide. The behavior of the order parameters at the CEP is shown in Fig.~\ref{fig: ord par}, whereas one illustrative plot in the spinodal region is provided in Fig.~\ref{fig: ord pars at mu=M}. We stress that these results are obtained with the conventional choice of parameters given above. In App.~\ref{app: params}, we investigate the dependence of the phase structure on the choice of parameters of the model.

That the pseudo-critical temperatures $T_{\rm pc}^\sigma$, $T_{\rm pc}^\ell$ and $T_{\rm pc}^{\bar\ell}$ all coincide at the CEP is not a surprise. This is because neither the chiral transformations nor the center transformations are exact symmetries in the presence of finite, non-zero quark masses and, thus, none of the quantities $\sigma$, $\ell$ or $\bar\ell$ is protected by a symmetry. Then, in particular, any discontinuity of one of the order parameters implies a discontinuity of the other ones. Since the CEP is located at the boundary of a first-order transition line, we deduce that the CEP is the same for the three order parameters. A more quantitative discussion of this issue is given in App.~\ref{app:CEP}, together with a proof that the spinodals also need to coincide.

\begin{figure}[t]
    \centering
    \includegraphics[width=0.9\linewidth]{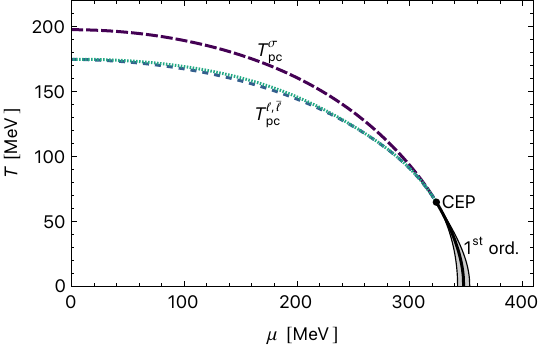}
    \caption{Phase diagram of the csCF-PNJL model in the $(\mu,T)$-plane for the parameters given in Sec.~\ref{sec: framework}. The dashed lines mark the crossovers of the various order parameters $\sigma$, $\ell$ and $\bar\ell$ (lighter, dotted), the black dot labels the critical endpoint, and the solid lines mark the spinodal region and first-order transition, equal for the three order parameters, see App.~\ref{app:CEP}.}
    \label{fig: standard phase dia}
\end{figure}

Finally, beyond $\smash{|\mu|=\mu_c(0)}$, $\sigma$, $\ell$ and $\bar\ell$ somewhat lose their role of order parameters in the sense that the extremal values as $\smash{T\to0}$ and $\smash{T\to\infty}$ are less far apart from each other than before, see Fig.~\ref{fig: ord par}. In fact, as we will see below, already beyond $\smash{|\mu|=M_0}$, the values of the order parameters at $\smash{T\to 0}$ undergo a considerable change from their constant values below $\smash{|\mu|=M_0}$. This is well known for the chiral condensate and relates to the breaking of the Silver-Blaze property at the first singularity on the $\mu$-axis of the phase diagram. For the Polyakov loops, however, most approaches lead to a Polyakov loop that vanishes in the limit $\smash{T\to 0}$ in some range beyond $\smash{|\mu|=M_0}$, which is usually interpreted as the system still being in a confined phase. The behavior seen in Fig.~\ref{fig: ord par} is a peculiarity of the center-symmetric CF potential and points, instead, to a deconfined phase. By scrutinizing other observables, we will argue below that, even in those models where the Polyakov loops vanish beyond $\smash{|\mu|=M_0}$, this could/should be interpreted as a deconfined phase.

The phase diagram of Fig.~\ref{fig: standard phase dia} summarizes the various features that we have discussed so far regarding the description of the transition in terms of the quantities $\sigma$, $\ell$, and $\bar\ell$. We recall that, in the crossover region, the transition temperatures are obtained from the resolution of the equations (\ref{eq:Tpc}). These equations may admit additional solutions. Although one should not necessarily associate them with crossover transitions as they do not mark a change between two extremal and different enough values of the corresponding order parameter, it is interesting to monitor the location of these inflections as they can reveal interesting connections between special points of the phase diagram. We perform a thorough analysis of the inflection points of $\sigma$, $\ell$, and $\bar\ell$ in the App.~\ref{app:inflections}.

Let us also stress that the $\mu$-axis of the phase diagram (corresponding to the $\smash{T\to 0}$ limit) reveals a number of interesting features, including additional singularities aside from the CEP. We postpone their analysis to the next section since they are intimately related to another interesting probe of the phase diagram, the net quark number response of the medium to an external quark probe. Additionally, the analysis of the various inflections of this quantity will reveal interesting connections between these singularities and the CEP.

\section{Zero-temperature limit and Net quark number response}\label{sec:T0}

When probing the different phases of QCD, it is interesting to look for observables that could directly probe the nature of the degrees of freedom that dominate in each phase. In recent works \cite{MariSurkau:2025MesBar, MariSurkau:2025XQCD, MariSurkau:2025suN}, we have argued that one possible such observable is the net quark number response of the medium to an external quark or antiquark probe. Since the relevance of these quantities is most clearly seen in the zero-temperature limit, we shall consider this limit first before discussing how they can be used to chart the phase diagram.

Actually, we shall first review the zero-temperature limits of the more conventional quantities $M$, $\ell$, and $\bar\ell$. If the discussion is pretty standard in the case of $M$, we argue that $\ell$ and $\bar\ell$ lead to some interpretational issues that we try to clarify, in particular, by contrasting them with the behavior of the net quark number responses.

\subsection{Preliminary remarks}
Preparing the ground for the analysis, we notice that the potential (\ref{eq: VPNJL}) rewrites identically as a function of $\smash{M=m_l+\sigma}$ and $(\ell,\bar\ell)$ as given in Eqs.~(\ref{eq:l0})-(\ref{eq:lb0}). One finds
\begin{widetext}
\beq\label{eq: V_q ells}
V_{\rm PNJL}(M,\ell,\bar\ell) & = &  \frac{(M-m_l)^2}{2G}-\frac{N_f}{\pi^2}\!\int_0^\Lambda dq\,q^2\Bigg\{3\varepsilon_q+T\ln\Big[1\!+\!3\ell e^{-\beta(\varepsilon_q-\mu)}\!+\!3\bar\ell e^{-2\beta(\varepsilon_q-\mu)}\!+\!e^{-3\beta(\varepsilon_q-\mu)}\Big]\nonumber\\
    & & \hspace{5.0cm}+\,T\ln\Big[1\!+\!3\bar\ell e^{-\beta(\varepsilon_q+\mu)}\!+\!3\ell e^{-2\beta(\varepsilon_q+\mu)}\!+\!e^{-3\beta(\varepsilon_q+\mu)}\Big]\,\Bigg\}\,,\label{eq:Vf}
\eeq
\end{widetext}
where we have introduced $\smash{\varepsilon_q\equiv\sqrt{q^2+M^2}}$. The exponentials appearing in the two logarithms are rapidly suppressed as $\smash{q\to\infty}$ over scales of the order of $T$. This means that, as long as $T$ lies enough below $\Lambda$ (which is anyway the situation in which the model makes sense), it should be possible to send $\Lambda$ to infinity for those terms, without affecting the results significantly. Above and below, our results are obtained using this prescription, as commonly done in the literature \cite{Ratti:2005jh, Fukushima:2008wg}, and motivated by the appearance of unphysical cutoff artifacts otherwise. While, for example, in our case the location of the critical endpoint is barely affected if we keep the cutoff, the behavior of the Polyakov loops changes somewhat at higher $\mu$, where non-monotonicities appear at finite $T$, and the crossovers are shifted and do not connect the critical endpoints at positive and negative chemical potentials. Cutoff artifacts also become more visible in thermodynamic quantities that derive from the potential, which can miss important contributions if the cutoff were implemented in the thermal integrals, and consequently not consistently approach the Stefan-Boltzmann limit at larger $T$. The behaviors at small enough temperatures and chemical potentials are however very similar with or without a thermal cutoff. Overall, in the simple two-flavor NJL model without diquarks considered here, the cutoff artifacts can still be controlled as the medium and vacuum contributions are easily separated. However, when considering color-superconductivity it becomes harder to separate medium and vacuum contributions and cutoff artifacts have to be dealt with more consistently, see e.g. \cite{Gholami:2024diy}.

We stress that the same exponentials are not always suppressed in the $\smash{T\to 0}$ limit. More precisely, when $\smash{|\mu|\geq M}$ there is a range of values of $q$ where some of the exponentials grow instead, thus affecting the way the corresponding integrals behave as $\smash{T\to 0}$.\footnote{This remains true for $\smash{|\mu|=M}$ where this range is limited to $\smash{q=0}$.} To cope with this efficiently, it is convenient to rewrite the additive term $3\varepsilon_q$ in Eq.~(\ref{eq:Vf}) as $3(\varepsilon_q-\mu)/2+3(\varepsilon_q+\mu)/2$ and to notice that, upon factorizing the appropriate exponential factors from arguments of the logarithms, one can flip the overall sign of each occurrence of $\varepsilon_q-\mu$ or $\varepsilon_q+\mu$ while exchanging the role of $\ell$ and $\bar\ell$. This means that one can rewrite the potential as
\begin{widetext}
\beq
V_{\rm PNJL}(M,\ell,\bar\ell) & = &  \frac{(M-m_l)^2}{2G}
   -\frac{3}{2}\frac{N_f}{\pi^2}\!\int_0^\Lambda dq\,q^2\Big\{|\varepsilon_q-\mu|+|\varepsilon_q+\mu|\Big\}\nonumber\\
    & & -\frac{TN_f}{\pi^2}\!\int_0^\infty dq\,q^2\Bigg\{\ln\Big[1\!+\!3\ell\star e^{-\beta|\varepsilon_q-\mu|}\!+\!3\bar\ell\star e^{-2\beta|\varepsilon_q-\mu|}\!+\!e^{-3\beta|\varepsilon_q-\mu|}\Big]\nonumber\\
    & & \hspace{3.0cm}+\,\ln\Big[1\!+\!3\bar\ell\star e^{-\beta|\varepsilon_q+\mu|}\!+\!3\ell \star e^{-2\beta|\varepsilon_q+\mu|}\!+\!e^{-3\beta|\varepsilon_q+\mu|}\Big]\,\Bigg\}\,,
\eeq
where we already took the liberty to send $\Lambda$ to $\infty$ in the exponentially convergent terms, and where $\star$ denotes an operator that keeps the Polyakov loop appearing on its left as it is when the absolute value of the argument in the exponential factor on its right is not needed, and replaces it by the charge-conjugated Polyakov loop otherwise. Although a bit formal, this notation will allow us to highlight the features of the $\smash{T\to 0}$ limit in a simple way. It will also be convenient to consider the changes of variables $\smash{x=\beta(\varepsilon_q-\mu)}$ and $\smash{x=\beta(\varepsilon_q+\mu)}$ which lead to
\beq
V_{\rm PNJL}(M,\ell,\bar\ell) & = &  \frac{(M-m_l)^2}{2G}
   -\frac{3}{2}\frac{N_f}{\pi^2}\!\int_0^\Lambda dq\,q^2\Big\{|\varepsilon_q-\mu|+|\varepsilon_q+\mu|\Big\}\nonumber\\
    & & -\frac{T^2N_f}{\pi^2}\!\int_{\beta(M-\mu)}^\infty dx\,(Tx+\mu)\sqrt{(Tx+\mu)^2-M^2}\ln\Big[1\!+\!3\ell\star e^{-|x|}\!+\!3\bar\ell\star e^{-2|x|}\!+\!e^{-3|x|}\Big]\nonumber\\
        & & -\frac{TN_f}{\pi^2}\!\int_{\beta(M+\mu)}^\infty dx\,(Tx-\mu)\sqrt{(Tx-\mu)^2-M^2}\ln\Big[1\!+\!3\bar\ell\star e^{-|x|}\!+\!3\ell \star e^{-2|x|}\!+\!e^{-3|x|}\Big]\,.\label{eq:pot1}
\eeq
\end{widetext}

As for $V_{\rm csCF}(r_3,r_8)$, it is not true that it rewrites identically in terms of $\ell$ and $\bar\ell$, but, over the appropriate domains of $r_3$ and $r_8$, we can invert the relations (\ref{eq:l0})-(\ref{eq:lb0}) to construct a function $V_{\rm csCF}(\ell,\bar\ell)$, see Ref.~\cite{MariSurkau:2025MesBar}. In fact, rather than discussing the particular case of $V_{\rm csCF}(\ell,\bar\ell)$, and following the logic of Ref.~\cite{MariSurkau:2025MesBar}, we will consider in what follows a generic class of model functions $V_{\rm glue}(\ell,\bar\ell)$ for the glue contribution to the potential. We require these model functions to be center-symmetric:
\beq
V_{\rm glue}(\ell,\bar\ell)=V_{\rm glue}(e^{i2\pi/3}\ell,e^{-i2\pi/3}\bar\ell)\,,\label{eq:center}
\eeq
and confining at low temperatures. By this, we mean the relevant extremum in this limit is located at $\smash{\ell=\bar\ell=0}$. Those are well-established properties of the glue sector. In addition, we assume that the glue contribution dominates over the quark contribution in the limit $\smash{T\to 0}$ and as long as $\smash{|\mu|<M}$. This is not a rigorously established property but a common feature of most, if not all, successful descriptions of the confinement/deconfinement transition. All the above properties are obeyed by $V_{\rm csCF}(\ell,\bar\ell)$ as well as by the various model glue potentials of Ref.~\cite{Fukushima:2003fw,Ratti:2005jh,Roessner:2006xn,Fukushima:2008wg,Lo:2013hla,Pisarski:2016ixt,Sasaki:2012bi,Canfora:2015yia,Kroff:2018ncl}. 

Let us now study the fate of the potential 
\beq
V(M,\ell,\bar\ell)\equiv V_{\rm glue}(\ell,\bar\ell)+V_{\rm PNJL}(M,\ell,\bar\ell)\,,
\eeq
and of its extrema as $\smash{T\to 0}$. The values of $M$, $\ell$ and $\bar\ell$ at the relevant extremum (the one from which one eventually extracts the physics) will be denoted $M(T,\mu)$, $\ell(T,\mu)$ and $\bar\ell(T,\mu)$, while their respective $\smash{T\to 0}$ limits will be denoted $M(\mu)$, $\ell(\mu)$ and $\bar\ell(\mu)$. We shall start by recalling known results for $M(\mu)$ before discussing the less studied case of $\ell(\mu)$ and $\bar\ell(\mu)$, as well as related quantities such as the net quark number responses $\Delta Q_q(\mu)$ and $\Delta Q_{\bar q}(\mu)$.

\subsection{Zero-temperature limit of \texorpdfstring{$M(T,\mu)$}{M(T,mu)} and Silver-Blaze property}\label{subsec: M(T to 0)}
\begin{figure}[t]
    \centering
    \includegraphics[width=\linewidth]{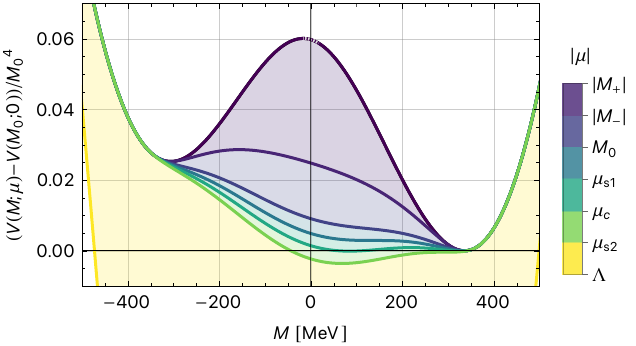}
    \caption{The zero-temperature potential $V(M;\mu)-V(M_0;0)$ normalized by $M_0^4$ for increasing values of $|\mu|$. The plain curves of decreasing darkness correspond to the potential at the special values of $\smash{|\mu|\in\{|M_+|,|M_-|,M_0,\mu_{s1},\mu_c,\mu_{s2},\Lambda\}}$, see the main text for the definitions, and the dotted, light gray line between $(-M_+,M_+)$ shows the potential at $\smash{\mu=0}$ in the range where it deviates (slightly) from the one at $M_+$.}
    \label{fig:VofM}
\end{figure}

When considering the behavior of the potential (\ref{eq:pot1}) as $\smash{T\to 0}$, we need to distinguish three regions, corresponding to $\smash{M<|\mu|}$, $\smash{M=|\mu|}$ and $\smash{M>|\mu|}$. As we will see below, in any case, the contribution to (\ref{eq:pot1}) that couples $M$, $\ell$, and $\bar\ell$ together is suppressed with respect to the contribution that depends only on $M$. It follows that, as far as the determination of $M(\mu)$ is concerned, the relevant potential is
\beq\label{eq: M potential T to 0}
V(M;\mu)\simeq\frac{(M-m_l)^2}{2G}\!-\!\frac{3N_f}{2\pi^2}\!\!\int_0^\Lambda \!\!\!\!\! dq\,q^2\Big\{|\varepsilon_q-\mu|\!+\!|\varepsilon_q+\mu|\Big\}.\nonumber\\
\eeq
It is useful to notice that, for any given $|\mu|$, the value of this potential  for $\smash{|M|\geq |\mu|}$ does not depend on $|\mu|$ and coincides with that of the potential at $\smash{\mu=0}$:
\beq
V(M;0)\simeq\frac{(M-m_l)^2}{2G}-\frac{3N_f}{\pi^2}\!\!\int_0^\Lambda \!\!\! dq\,q^2\,\varepsilon_q\,.
\eeq
For the considered parameters, the latter potential has three extrema: a global minimum which occurs by definition at $\smash{M=M_0\simeq 336}$\,MeV, together with a local minimum at $\smash{M=M_-\simeq -306}$\,MeV and a local maximum at $\smash{M=M_+\simeq -17}$\,MeV, see the upper curve shown in Fig.~\ref{fig:VofM}. We refer to App.~\ref{app:VofM} for a thorough study of the possible shapes of this vacuum potential depending on the parameters.

As $\mu$ is slightly taken away from $0$, this potential is modified in the tiny region $\smash{|M|<|\mu|}$. Then, for small enough $|\mu|$, the potential has exactly the same extrema as the potential at $\smash{\mu=0}$. For finite $|\mu|$, what happens depends on the shape of the potential in the inner region $\smash{|M|<\mu}$ but it should be clear that any extremum $M_{\rm extr}$ of the potential at $\smash{\mu=0}$ should remain an extremum of the potential at finite $\mu$ as long as $|\mu|<|M_{\rm extr}|$. In other words, if we follow continuously the position of any extremum $M_{\rm extr}$ existing already at $\smash{\mu=0}$, we should have
\beq
M_{\rm extr}(\mu)=M_{\rm extr}\,, \quad \mbox{for } |\mu|<|M_{\rm extr}|\,.
\eeq
We illustrate this feature in Fig.~\ref{fig:VofM} where we show the evolution of the zero-temperature potential (for the parameters considered in this work) as $|\mu|$ is increased.

Of course, the relevant extremum is the global minimum, and even though it corresponds to $M_0$ for low enough $\mu$, it could happen that it does not remain equal to $M_0$ all the way up to $\smash{|\mu|=M_0}$. This is because new extrema might develop in the inner region and become the absolute minima for some value $\smash{|\mu|=\tilde M_0}$, with $\smash{0<\tilde M_0<M_0}$. We can still claim, however, that\footnote{Since the thermal corrections to the zero-temperature potential (\ref{eq: M potential T to 0}) are suppressed by the inverse temperature exponentials and factors of Polyakov loops, it is reasonable to expect this solution to hold approximately even at temperatures below $M_0$ and as long as the Polyakov loops are small. A crude estimate may be ${\left.T\right|_{M\simeq M_0}(\mu)\lesssim(M_0^2-\mu^2)/2M_0}$, which we find to work quite well for the current model.}
\beq
M(\mu)=M_0\,, \quad \mbox{for } |\mu|<\tilde M_0\,.
\eeq
This result is well-known and a particular illustration of the so-called Silver-Blaze property \cite{Cohen:2003kd}, which arises from a combination of analyticity and the effect of certain U(1) transformations that do not change the boundary conditions of the quark fields in the Euclidean functional integral formulation of QCD \cite{Marko:2014hea}. 

More precisely, the Abelian transformations imply that 
\beq
M(T,\mu)=M(T,\mu+i2\pi n T)\,,
\eeq
with $\smash{n\in\mathds{Z}}$. Choosing $\smash{T=\Delta\mu_i/(2\pi n)}$ for a given $n$ and a given $\smash{\Delta\mu_i\in\mathds{R}}$, one finds
\beq
M\left(\frac{\Delta\mu_i}{2\pi n},\mu\right)=M\left(\frac{\Delta\mu_i}{2\pi n},\mu+i\Delta\mu_i\right).
\eeq
Sending $n$ to infinity at fixed $\Delta\mu_i$, this becomes
\beq
M(\mu)=M(\mu+i\Delta\mu_i)\,,
\eeq
which shows that $M(\mu)$, seen as a function of a complex $\mu$, depends only on the real part of $\mu$:
\beq
M(\mu)=M({\rm Re}\,\mu)\,.\label{eq:per}
\eeq
Now, if $M(\mu)$ is analytic in a neighborhood of $\smash{\mu=0}$, owing to Eq.~(\ref{eq:per}), it has to be analytic over an open band containing the axis $\smash{{\rm Re}\,\mu=0}$. But because it is constant along this axis, it has to be constant along the whole band, and, thus, in particular, over the real axis, in some neighborhood of $\smash{\mu=0}$.

This will be true until a first singularity is reached for a given value of ${\rm Re}\,\mu$. After this point, although one still has (\ref{eq:per}), the function $M(\mu)$ has no reason to be analytic, and thus it has no reason to be constant over the real axis.\footnote{Below, we will see some examples of quantities which are again analytic in some band after this first singularity and which are then again constant along the real axis, but with a different constant value than before the singularity.} In fact, because the function remains constant along the imaginary direction according to (\ref{eq:per}), the non-analyticity is tantamount to the fact that the function is not constant along the real direction. For the parameters considered in this work, it seems that no new extrema appear in the inner region $\smash{|M|<|\mu|}$,\footnote{Note, however, that the two other extrema for $\smash{M<0}$ start varying with $\mu$ respectively beyond $\smash{|\mu|=|M_+|}$ and $\smash{|\mu|=|M_-|}$ and disappear at values of $|\mu|$ below $M_0$. Note also that, for certain parameters, type a) of Sec.~2.3.2 in Ref.~\cite{Buballa:2003qv}, new extrema can appear in the inner region, and a first order transition can occur at $\smash{|\mu|=\Tilde{M}_0<M_0}$, with a jump directly into a chirally restored $\smash{M(|\mu|>\Tilde{M_0})\ll M_0}$.} for $\smash{|\mu|<M_0}$ and thus we have $\smash{\tilde M_0=M_0}$.

Beyond $\smash{|\mu|=M_0}$, $M(\mu)$ decreases continuously, though not analytically, away from $M_0$. As we will see in Sec.~\ref{sec: thermodynamics}, see in particular Fig.~\ref{fig: nq-norm}, the zero-temperature net quark number density $n_q$ goes continuously from zero to non-zero at this point, which is then reminiscent of the nuclear liquid-gas transition. In the present PNJL model, however, this transition is continuous at $\smash{T=0}$ and becomes a smooth crossover for finite temperatures, see below. This is a shortcoming of the PNJL model at finite densities since one expects, instead, a first-order type transition along the $\mu$-axis in the QCD case \cite{Elliott:2013pna}. Let us stress that the model does not directly account for baryons.\footnote{See, however, the discussion below for some indirect account of the baryons.} In particular, the onset of the dense phase is not shifted down by the nuclear binding energy, but appears directly at the quark mass $|\mu|\simeq 336\,$MeV in the Silver-Blaze region.

Going towards higher $|\mu|$, for our choice of parameters, we find a spinodal region $\smash{\mu_{s1}(0)<|\mu|<\mu_{s2}(0)}$, with $\smash{\mu_{s1}(0)\simeq 343\,}$MeV and $\smash{\mu_{s2}(0)\simeq 353\,}$MeV, where the $\smash{T=0}$ gap equation actually has three (positive mass) solutions (as already mentioned above, the solutions with negative $M$ have already ceased to exist below $\smash{|\mu|=M_0}$). The two spinodals lie on each side of a first-order transition occurring at ${\mu_c(0)\simeq348\,}$ MeV, which connects to the CEP found in the $(\mu,T)$-plane. After the first-order transition, the chiral condensate becomes quite small, and $M$ is not much larger than $m_l$. Thus, with the used parameters, the model is of the type b), as defined in Sec.~2.3.2 of Ref.~\cite{Buballa:2003qv}, with first a smooth decrease and then a first-order jump of the quark mass.

\subsection{Zero-temperature limit of \texorpdfstring{$\ell(T,\mu)$ and $\bar\ell(T,\mu)$}{the Polyakov loops}}

Let us proceed to a similar discussion for the other order parameters at our disposal, the Polyakov loops $\ell(\mu)$ and $\bar\ell(\mu)$. For simplicity, we assume that the parameters are such that $\smash{\tilde M_0=M_0}$, which is the case for the presently considered parameters.

Suppose first that $\smash{|\mu|<M_0}$. We have seen in this case that $\smash{M(\mu)=M_0}$, and, thus, as far as the determination of $\ell(\mu)$ and $\bar\ell(\mu)$ is concerned, we can set $\smash{M=M_0}$ in the potential. Since $\smash{|\mu|<M_0}$, the lower integration boundaries  $\smash{\beta(M_0-\mu)}$ and $\smash{\beta(M_0+\mu)}$ in Eq.~(\ref{eq:pot1}) both approach $+\infty$ as $\smash{T\to 0}$. Then, the corresponding integrals are exponentially suppressed due to the exponential factors inside the logarithms. Since we have assumed that $V_{\rm glue}(\ell,\bar\ell)$ is not as suppressed in this limit, the relevant potential for the determination of $\ell(\mu)$ and $\bar\ell(\mu)$ is simply $V_{\rm glue}(\ell,\bar\ell)$, and, because we have assumed that the latter is confining at low temperatures, we find that
\beq
\ell(\mu)=\bar\ell(\mu)=0\,, \quad \mbox{for } |\mu|< M_0\,.\label{eq:SBl}
\eeq
This is yet another illustration of the Silver-Blaze property but we note that we have not yet checked whether $M_0$ is the maximal value of $|\mu|$ for which the two Polyakov loops vanish. As we will see, this depends on the considered model for the glue potential. Yet, we will argue that the interpretation of the phase just above $\smash{|\mu|=M_0}$ is independent of the chosen glue potential.

Suppose now that $\smash{|\mu|>M_0}$. In this case, we have seen that $M(\mu)$ is not constant anymore, but, as far as the determination of $\ell(\mu)$ and $\bar\ell(\mu)$ is concerned, we can set $\smash{M=M(\mu)}$ in the potential. Since $\smash{M(\mu)<M_0}$, we deduce that $\smash{M(\mu)<|\mu|}$, and then, while one of the two boundaries $\beta(M(\mu)-\mu)$ and $\beta(M(\mu)+\mu)$ still approaches $+\infty$ as $\smash{T\to 0}$, the other one approaches $-\infty$. It follows that, while one of the integrals is still exponentially suppressed in this limit, the other one is only polynomially suppressed. 
Suppose for instance that $\smash{\mu>M_0}$. Then, the relevant potential for determination of $\ell(\mu)$ and $\bar\ell(\mu)$ is
\beq
& & V(\ell,\bar\ell)\simeq V_{\rm glue}(\ell,\bar\ell)-\frac{T^2N_f}{\pi^2}\mu\sqrt{\mu^2-M^2(\mu)}\nonumber\\
& & \hspace{0.5cm}\times\,\int_{-\infty}^\infty dx\,\ln\Big[1\!+\!3\ell\star e^{-|x|}\!+\!3\bar\ell\star e^{-2|x|}\!+\!e^{-3|x|}\Big]\,.\nonumber\\
\eeq
Making use of the definition of $\star$ while splitting the $x$-integral over $\smash{x>0}$ and $\smash{x<0}$, and upon making the change of variables $\smash{x\to -x}$ in the latter, this rewrites
\beq
& & V(\ell,\bar\ell)\simeq V_{\rm glue}(\ell,\bar\ell)-\frac{T^2N_f}{\pi^2}\mu\sqrt{\mu^2-M^2(\mu)}\nonumber\\
& & \hspace{0.5cm}\times\int_0^\infty dx\,\ln\Big[1\!+\!3\ell e^{-x}\!+\!3\bar\ell e^{-2x}\!+\!e^{-3x}\Big]\nonumber\\
& & \hspace{2.5cm}\times\Big[1\!+\!3\bar\ell e^{-x}\!+\!3\ell e^{-2x}\!+\!e^{-3x}\Big]\,.
\eeq
If $\smash{\mu<-M_0}$, we find a similar expression with $\mu\to -\mu$, which amounts to replacing $\mu$ by $|\mu|$ in the previous formula. Notice that the above potential is invariant under $\ell\leftrightarrow\bar\ell$. If we assume that, for the chosen $\smash{M=M(\mu)}$, there is only one extremum (this is what we find numerically), we deduce that $\smash{\ell(\mu)=\bar\ell(\mu)}$. This also means that in order to access $\ell(\mu)$, we can extremize the reduced potential
\beq
V(\ell) & = & V_{\rm glue}(\ell,\ell)-\frac{T^2N_f}{\pi^2}|\mu|\sqrt{\mu^2-M^2(\mu)}\nonumber\\
& & \times\int_0^\infty dx\,\ln\Big[1\!+\!3\ell (e^{-x}\!+\! e^{-2x})\!+\!e^{-3x}\Big]^2.\label{eq: gap ell T=0}
\eeq

The exact behavior of the Polyakov loops depends then on whether and/or how much the temperature scaling of $V_{\rm glue}(\ell,\bar\ell)$ is suppressed compared to the $T^2$-correction in Eq.~(\ref{eq: gap ell T=0}). 

For models such as the ones used in Refs.~\cite{Fukushima:2003fw, Ratti:2005jh, Roessner:2006xn, Fukushima:2008wg}, it is less suppressed and the Polyakov loops still vanish for $\smash{T\to 0}$ extending the Silver-Blaze property (\ref{eq:SBl}) beyond $\smash{|\mu|=M_0}$. For the present model based on the center-symmetric CF potential (\ref{eq: VcsLg}), and the one used in Refs.~\cite{Pisarski:2016ixt, Folkestad:2018psc}, the glue contribution in Eq.~(\ref{eq: gap ell T=0}) scales exactly as $T^2$, and therefore the Polyakov loops can acquire a finite value, making the Silver-Blaze property (\ref{eq:SBl}) stop at $\smash{|\mu|=M_0}$, in line with the critical behavior of $M(\mu)$ at the same point. Due to the presence of the factor $\sqrt{\mu^2-M^2(\mu)}$ in Eq.~(\ref{eq: gap ell T=0}) and the fact that $V_{\rm glue}(\ell,\ell)$ is confining at low temperatures, the loops approach $0$ as $\smash{|\mu|\to M_0^+}$, so that $\ell(\mu)$ and $\bar\ell(\mu)$ are continuous at $\smash{|\mu|=M_0}$. In models such as the one used in Refs.~\cite{Reinosa:2014ooa, Canfora:2015yia,Kroff:2018ncl}, where the gauge sector is suppressed more strongly, the Polyakov loops can instead jump to a finite value at $\smash{|\mu|=M_0}$. Note also that, in the present model, since the Polyakov loops are coupled to $M$ via the second term of Eq.~\eqref{eq: gap ell T=0}, they experience the same spinodal region and first-order transition. In other models, this transition may be suppressed as a consequence of the dominance of the glue potential over the $\sim T^2$ quark contribution.\footnote{One may wonder how this fits the discussion in App.~\ref{app:CEP} where we argue that features such as critical endpoints, first order transitions, or spinodals are the same for $M$, $\ell$, $\bar\ell$. This discussion, however, applies in a context where both chiral symmetry and center symmetry are explicitly broken, while in the $\smash{T\to 0}$ limit, center symmetry can be restored due to the fact that the quark boundary condition that usually break the symmetry can become irrelevant.}

These rather different behaviors of the Polyakov loops for $\smash{|\mu|>M_0}$ and $\smash{T\to 0}$, depending on the considered model for the glue sector, seem to pose an interpretation problem on the nature of the phase beyond $\smash{|\mu|=M_0}$. In the case where the Polyakov loops vanish, as in the models of Refs.~\cite{Fukushima:2003fw, Ratti:2005jh, Roessner:2006xn, Fukushima:2008wg}, it is tempting to interpret this as the signal of a confined phase, whereas for other models, such as those of Refs.~\cite{vanEgmond:2021jyx, Reinosa:2014ooa, Pisarski:2016ixt, Canfora:2015yia} where the Polyakov loops do not vanish, one should instead consider that the system is in a deconfined phase. What we will argue in the next section is that, irrespective of the different behaviors for the Polyakov loops, the interpretation of the region $\smash{|\mu|>M_0}$ should be that of a deconfined phase.\footnote{Our model does not include the possibility of color superconducting phases and Cooper pairing.} This may sound a little unorthodox since we are used to the idea that a vanishing Polyakov loop corresponds to a confined phase. We will argue below, however, that the Polyakov loop is not the most relevant observable in this limit and that other observables allow for a cleaner interpretation. For the time being, we only stress that even when the Polyakov loops vanish as $\smash{T\to 0}$ for $\smash{|\mu|>M_0}$, they do so in a much slower (polynomial) way than the exponential suppression obtained for $\smash{|\mu|<M_0}$. This is already an indication that the two regions are actually rather different.

\subsection{Net quark number responses \texorpdfstring{$\Delta Q_q$ and $\Delta Q_{\bar q}$}{DeltaQq and DeltaQqbar}}

We have seen that, as long as $\smash{|\mu|<M_0}$, the Polyakov loops vanish as $\smash{T\to 0}$, and this, no matter what model is considered. It turns out that determining how fast the Polyakov loops vanish with $T$ reveals much more information.

To this purpose, we expand the potential to quadratic order in $\ell$ and $\bar\ell$, around $\smash{\ell=\bar\ell=0}$, while setting $\smash{M=M_0}$ in the $\ell$- and $\bar\ell$-dependent parts. For this endeavour, it is enough and convenient to work with the expression (\ref{eq:Vf}). Moreover using the center symmetry (\ref{eq:center}), it is easily checked that $\partial_\ell V_{\rm glue}$, $\partial_{\bar\ell} V_{\rm glue}$, $\partial_{\ell}^2V_{\rm glue}$ and $\partial_{\bar\ell}^2V_{\rm glue}$ vanish at $\smash{\ell=\bar\ell=0}$. Then, the relevant potential for obtaining $\ell(T,\mu)$ and $\bar\ell(T,\mu)$ at low $T$ is
\beq
V(\ell,\bar\ell) & \simeq  & \partial_\ell\partial_{\bar\ell}V_{\rm glue}\times \ell\bar\ell\nonumber\\
& - & C(e^{\beta\mu}f_{\beta M_0}+e^{-2\beta\mu}f_{2\beta M_0})\ell\nonumber\\
& - & C(e^{-\beta\mu}f_{\beta M_0}+e^{2\beta\mu}f_{2\beta M_0})\bar\ell\,,\label{eq:V1}
\eeq
where $\smash{C=3N_fTM^3_0}$ and we have introduced the function
\begin{equation}
    f_y\equiv \frac{1}{\pi^2}\int_0^\infty \!\!\! dx\,x^2e^{-y\sqrt{x^2+1}}\sim \frac{y^{-3/2}}{\sqrt{2}\pi^{3/2}}e^{-y}\,,\label{eq: f_y}
\end{equation}
with the asymptotic expression on the right holding for $y\to\infty$. Note that we have omitted quadratic contributions in $\ell$ and $\bar\ell$ coming from the logarithms in Eq.~(\ref{eq:Vf}) since those lead to subleading contributions to the leading behavior $\ell(T,\mu)$ and $\bar\ell(T,\mu)$ in the $\smash{T\to 0}$ limit. 
\begin{figure}[t]
    \centering
    \includegraphics[width=0.9\linewidth]{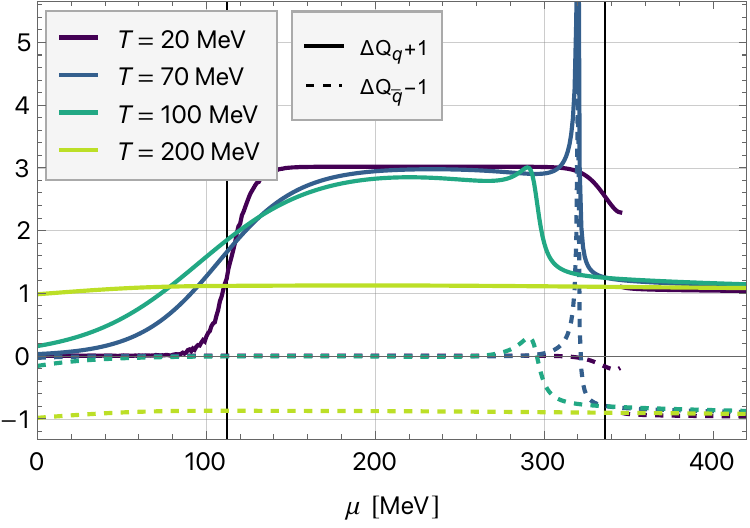}
    \caption{Net quark number gains as a function of the chemical potential $\mu$ at various temperatures $T$. Plain and dashed curves correspond respectively to the case of a quark ($\Delta Q_q$) or antiquark ($\Delta Q_{\bar q}$) probe. The black vertical lines mark the points $\mu=M_0/3$ and $\mu=M_0$.}
    \label{fig: DeltaQ}
\end{figure}
The extremization of (\ref{eq:V1}) with respect to $\ell$ and $\bar\ell$ then leads to
\beq
    \ell & \simeq & \frac{C}{\partial_\ell\partial_{\bar\ell} V_{\rm glue}}\left(e^{-\beta\mu} f_{\beta M_0}+e^{2\beta\mu} f_{2\beta M_0}\right), \label{eq: l Tto0}\\
    \bar\ell & \simeq & \frac{C}{\partial_\ell\partial_{\bar\ell} V_{\rm glue}}\left(e^{\beta\mu} f_{\beta M_0}+e^{-2\beta\mu} f_{2\beta M_0}\right), \label{eq: lb Tto0}
\eeq
which control how $\ell(T,\mu)$ and $\bar\ell(T,\mu)$ approach $0$ as $\smash{T\to 0}$ for $\smash{|\mu|<M_0}$.

As described in Refs.~\cite{MariSurkau:2025MesBar, MariSurkau:2025XQCD}, the physical content of (\ref{eq: l Tto0})-(\ref{eq: lb Tto0}) is captured by considering the related quantities
\beq
\Delta Q_q\equiv T\partial_\mu\ln\ell \quad \mbox{and} \quad \Delta Q_{\bar q}\equiv T\partial_\mu\ln\bar\ell\,,
\eeq
which give access to the net quark number response of the medium when a quark or an antiquark is added, respectively. By adding or subtracting $1$ to these quantities, one obtains the net quark numbers gained by the system, including the quark or antiquark probe.  From Eqs.~(\ref{eq: l Tto0})-(\ref{eq: lb Tto0}), it can be seen that these quantities follow a particular behavior at low temperatures, reminiscent of the absorption of the probe into states with the same net quark number as mesons or baryons. More precisely, a quark (resp. antiquark) probe is always absorbed into a meson-like state or a baryon-(resp. antibaryon-) like state, corresponding to a net quark number gain of the system of either $0$ or $+3$ (resp. $-3$). For negative chemical potentials, there is an excess of antiquarks in the medium, and it is thus always energetically favorable for a quark probe to combine into a meson-like state using an antiquark of the medium. This remains true for positive chemical potential provided that the excess of quarks over antiquarks remains moderate, corresponding to $\smash{\mu<M_0/3}$. Beyond this value, the excess of quarks is such that it becomes more favorable to combine into a baryon-like state using two quarks of the medium. Similar (mirrored) conclusions are obtained in the case of an antiquark probe. We illustrate this discussion in Fig.~\ref{fig: DeltaQ} with one of the smallest temperatures we could test at all $\mu$.\footnote{At very small $\mu$ and $T$ the numerical integrals need higher precision computations to evaluate, but at larger $\mu$ they are more controlled and we can easily access tiny temperatures. We find that the plateau at 3 or 0 extend until $\smash{\mu=M_0}$, where they sharply transition to $\pm1$ for all $\mu>M_0$.} The above discussion extends the heavy-quark QCD results of Refs.~\cite{MariSurkau:2025MesBar, MariSurkau:2025XQCD} to realistic quark masses. Below, we will show how thermal fluctuations soften the distinction between the meson- and baryon-type screening, which is a negligible effect for heavy quarks but becomes important at physical quark masses, as can already be guessed from Fig.~\ref{fig: DeltaQ}.

In the limit $\smash{T\to 0}$, the functions $\Delta Q_q(\mu)$ and $\Delta Q_{\bar q}(\mu)$ become discontinuous at $\smash{\mu=M_0/3}$ and $\smash{\mu=-M_0/3}$ respectively, which places new singularities along the $\mu$-axis before the ones at $\smash{|\mu|=M_0}$. We note, however, that this does not jeopardize the Silver-Blaze property for $M(\mu)$, $\ell(\mu)$ or $\bar\ell(\mu)$ which applies until $\smash{|\mu|=M_0}$. The point is that $\smash{\mu=\pm M_0/3}$ are not singularities for $M(\mu)$, $\ell(\mu)$ or $\bar\ell(\mu)$ but are singularities, respectively, of the ratio of $\ell$ and $T\partial_\mu\ell$ defining $\Delta Q_q$ and of the ratio of $\bar\ell$ and $T\partial_\mu\bar\ell$ defining $\Delta Q_{\bar q}$. Since $\ell$ and $T\partial_\mu\ell$ both vanish in the $\smash{T\to 0}$ limit, their ratio can generate new singularities, as it is precisely the case at $\smash{\mu=M_0/3}$ where, in a certain sense, $\ell$ reaches $0$ first, and similarly for the ratio of $\bar\ell$ and $T\partial_\mu\bar\ell$ at $\smash{\mu=-M_0/3}$. This means that the constant values of $\Delta Q_q$ for $\smash{-M_0<\mu<M_0/3}$ and of $\Delta Q_{\bar q}$ for $\smash{-M_0/3<\mu<M_0}$ are again a consequence of the Silver-Blaze property. In fact, the other constant values taken by $\Delta Q_q$ above $\smash{\mu=M_0/3}$ and by $\Delta Q_{\bar q}$ below $\smash{\mu=-M_0/3}$ are also reflections of the Silver-Blaze principle and tell us that the functions $\Delta Q_q(\mu)$ and $\Delta Q_{\bar q}(\mu)$ are not only analytic in the bands $\smash{\{\mu\in\mathds{C}\,|\,-M_0<{\rm Re}\,\mu<M_0/3\}}$ and $\smash{\{\mu\in\mathds{C}\,|\,-M_0/3<{\rm Re}\,\mu<M_0\}}$, respectively, but also in the bands $\smash{\{\mu\in\mathds{C}\,|\,M_0/3<{\rm Re}\,\mu<M_0\}}$ and $\smash{\{\mu\in\mathds{C}\,|\,-M_0<{\rm Re}\,\mu<M_0/3\}}$, respectively, see the discussion in the previous section and in footnote 10.

We now would like to argue that $\Delta Q_q$ and $\Delta Q_{\bar q}$ are also constant past the second singularity at $\smash{|\mu|=M_0}$, respectively, with a constant value equal to $0$. In the case where the glue contribution dominates, the Polyakov loops vanish with a prefactor $T^2|\mu|\sqrt{\mu^2-M^2(\mu)}$ from which it follows that $\Delta Q_q$ and $\Delta Q_{\bar q}$ vanish linearly with $T$. Similarly, when the glue contribution is of the same order as the rest, the Polyakov loops approach some function of $\mu$, and then again $\Delta Q_q$ and $\Delta Q_{\bar q}$ vanish linearly with $T$. This means, in particular, that the singularity at $\smash{\mu=M_0/3}$ is connected to those at $\smash{|\mu|=M_0}$ by the same quantity, the net quark number response $\Delta Q_q$, and similarly for the singularity at $\smash{\mu=-M_0/3}$ and those at $\smash{|\mu|=M_0}$, via $\Delta Q_{\bar q}$. We will see below that this result is in fact deeper since these same quantities will connect these singularities along the $\mu$-axis to the CEP in the bulk of the phase diagram. 

But the vanishing of $\Delta Q_q$ and $\Delta Q_{\bar q}$ for $\smash{|\mu|>M_0}$ also allows one to bring some clarification regarding the nature of the phase beyond that singularity. Indeed, while the behavior of the Polyakov loops is model-dependent in this region, the fact that both $\Delta Q_q$ and $\Delta Q_{\bar q}$ vanish independently of the model indicates that the system is in a deconfined phase where the net quark number gained by the medium after bringing a probe is given by the net quark number of the probe itself. These conclusions can be corroborated by looking at the change in the medium free-energy before and after bringing the probe,
\beq
\Delta F_q=-T\ln\ell \quad \mbox{and} \quad \Delta F_{\bar q}=-T\ln\bar\ell\,,
\eeq
which are related to $\Delta Q_q$ and $\Delta Q_{\bar q}$ as
\beq
\Delta Q_q=-\partial_\mu \Delta F_q \quad \mbox{and} \quad \Delta Q_{\bar q}=-\partial_\mu \Delta F_{\bar q}\,.
\eeq
Independently of the considered glue model, for $\smash{-M_0<\mu<M_0/3}$ and $\smash{M_0/3<\mu<M_0}$, we find respectively that, 
\beq
\Delta F_q=M_0+\mu\,,\quad\mbox{or}\quad\Delta F_q=2(M_0-\mu)\,,
\eeq
which are both non-zero and indicate that bringing a quark into the medium has some energy cost. In contrast, for $|\mu|>M_0$, one finds instead that
\beq
\Delta F_q=0\,,
\eeq
indicating that a quark can be brought into the medium with no energy cost. Similar conclusions apply in the case of an antiquark probe.

We note that these conclusions, although independent of the considered glue contribution to the potential, are derived within a model for QCD. Therefore, we cannot confirm with certainty whether they would still apply in full QCD. It might seem counterintuitive that a deconfined phase appears below the chiral transition line. However, we should stress that having a deconfined phase in the sense described above does not mean that stable bound states of quarks cannot exist in this region (in fact, we expect baryons and/or diquarks), but rather that the strict confinement of quarks, as measured by their free-energy found through the Polyakov loops, disappears, and isolated quarks can exist.

We emphasize that it is crucial to distinguish two distinct notions of confinement here. The first being in the theoretical sense that at low temperatures isolated quark states are forbidden, related to the center symmetry of Yang-Mills theory, and the second being the existence of hadronic (or color-superconducting) bound states, into which quarks are ``confined" in practice, simply because they are stable, multiquark configurations. While the first surely implies the second, the inverse is not necessarily true. What we find in the PNJL model is that the first type of confinement ceases to exist after $|\mu|>M_0$, but at this point we cannot make a statement about the ``empirical" confinement into bound states, and what degrees of freedom (the ground state) dominate the thermodynamics. The PNJL model does not directly incorporate baryon bound states, and the only observation that can be made comes from looking at the Polyakov loop-modified quark Fermi-Dirac distribution functions, which are model-dependent.

We note that this also indicates that the PNJL model has no ``quarkyonic" phase in the sense of \cite{McLerran:2007qj}, since the high $\mu$ region is deconfined, in contrast to the expectation from the large $N_c$ limit. The appearance of a quarkyonic phase has commonly been assumed in the PNJL literature, as in popular models the Polyakov loops vanish for $\smash{T\to0}$. The above treatment shows that one needs to be careful in defining confinement.

\subsection{Alternative look at the phase diagram}

Let us summarize what we have found so far along the $\smash{T\to 0}$, $\mu$-axis of the phase diagram. The order parameters $M(\mu)$, $\ell(\mu)$ and $\bar\ell(\mu)$ are essentially constant and equal to their vacuum ($\smash{\mu=T=0}$) solutions in the region $\smash{|\mu|<M_0}$. Upon inspection, however, the Polyakov loops, through the associated net quark number gains, show the screening of quarks into meson- and baryon-like states, with a transition between the two screening modes at $\smash{\mu=M_0/3}$ which appears as a singularity of $\Delta Q_q$, and similarly for the screening of antiquarks at $\smash{\mu=-M_0/3}$.

\begin{figure}[t]
    \centering
    \includegraphics[width=0.9\linewidth]{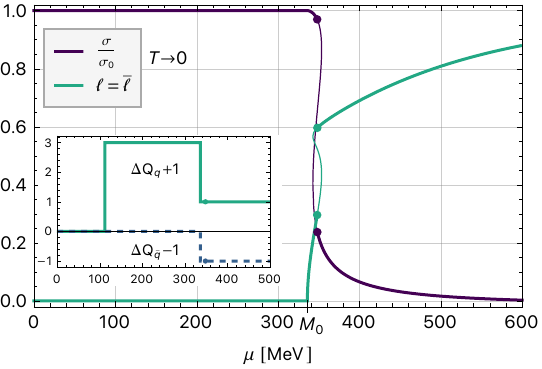}
    \caption{Order parameters as functions of $\mu$ at $\smash{T=0}$, showing the Silver-Blaze property. We inset the ${T\to0}$ limits of the net quark number gains. Note that, while $\smash{\ell=\bar\ell}$ for $\smash{T\to 0}$, their derivatives are not equal, hence the different net quark number gains.}
    \label{fig: silver-blaze}
\end{figure}

Then, at $\smash{|\mu|=M_0}$, there is another singularity where the constituent quark mass begins to decrease, and a dense phase sets in, similar to the nuclear liquid-gas transition but not with the expected order due to the lack of explicit baryons in the model. In the PNJL model, we find additionally that quark deconfinement sets in here, and the chiral condensate starts decreasing.\footnote{Note that this is a quantum transition, rather than a thermal one, and corresponds to a change in the ground state.} The net quark number responses are again pivotal in reaching this conclusion since the behavior of the Polyakov loops is model-dependent. In contrast to other models, however, in the center-symmetric CF model, the Polyakov loops also start acquiring non-zero values, which reinforces the interpretation that isolated quark and antiquark states are not forbidden beyond $\smash{|\mu|=M_0}$. They may still be suppressed (in the sense of their contribution to the thermodynamic distribution functions) by the small values of the Polyakov loops, making baryon-like three-quark states the dominant degrees of freedom. Finally, for slightly higher $|\mu|$, the first-order chiral transition occurs, to which a first-order jump of the Polyakov loops is related. After this, there is a chirally restored, deconfined phase, which continues as long as the model can be trusted ${|\mu|<\Lambda}$, with decreasing chiral condensate and increasing Polyakov loops. We show the behavior of the order parameters in Fig.~\ref{fig: silver-blaze}.

\begin{figure*}[t]
    \centering
    \includegraphics[width=0.9\linewidth]{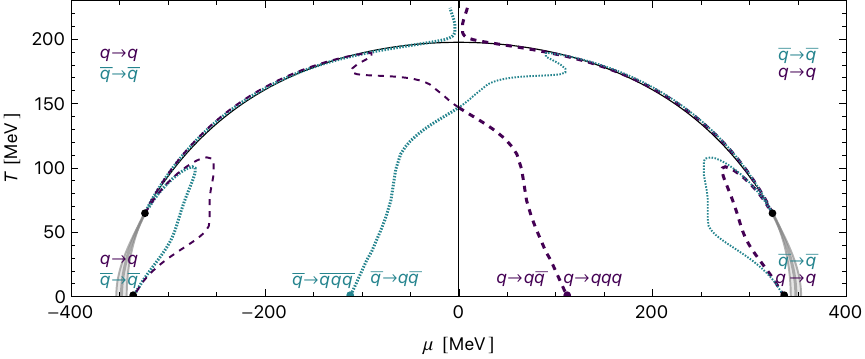}
    \caption{Phase diagram with pseudocritical temperature curves as obtained from the inflections of the net quark number gains ${\Delta Q_q+1}$ (dark, dashed) and ${\Delta Q_{\bar q}-1}$ (light, dotted). The thin black line shows the position of the chiral crossover, and the points at ${T=0}$ mark the jumps between 0 and $\pm3$ or between $\pm3$ and $\pm1$ for a quark/antiquark probe, and the divergences at the CEP. The notation $\smash{X\to Y}$ indicates that a probe $X$ brought to the medium is absorbed into a state $Y$.}
    \label{fig: phase dia from DQ}
\end{figure*}

This summary highlights the role of the net quark number responses in the interpretation of the various phases that occur along the $\mu$-axis. We now would like to argue that these same quantities allow one to chart the phase diagram in the $(\mu,T)$ plane while connecting the singularities along the $\mu$-axis to the CEP. Take, for instance, the singularity at $\smash{|\mu|=M_0}$. In the present model, the associated transition becomes a crossover at finite temperature, and thus the singularity turns into inflections of the various quantities at our disposal. Following these inflections can sometimes establish relations with other features of the phase diagram. In the present case, by studying the temperature dependence of the Polyakov loops in the vicinity of the singularity, we find that the corresponding inflections exhibit a closed-loop pattern that originates and ends at the point $\smash{(|\mu|,T)=(M_0,0)}$. The same is true for the (normalized) net quark number $n_q/T^3$. Instead, we find no inflection points for the chiral condensate at finite temperatures. The same exercise applied to the net quark number reveals more interesting patterns.

Let us first stress that, as the temperature is increased, we should expect the net quark number gains to correspond to a statistical combination of the various possibilities of recombination of the probe that we identified at zero temperature. This is indeed what we find since $\Delta Q_q+1$ lies predominantly between $0$ and $3$, while $\Delta Q_{\bar q}-1$ lies between $0$ and $-3$, see Fig.~\ref{fig: DeltaQ}. Rather than sharp transitions between $0$ and $\pm 3$, we now have a smooth crossover between a small value and a value close to $\pm 3$. This crossover can be identified by means of the inflections of $\Delta Q_q$ and $\Delta Q_{\bar q}$. As the temperature is increased further, above the deconfinement transition, and also above $\smash{|\mu|\simeq M_0}$, the net quark number gains become close to the net quark number of the probes, in line with the fact that, in the deconfined phase, the probe should have no sensible effect on the net quark number of the medium. Once again, the smooth transition between the confined phase and the deconfined phase can be identified via the inflections of $\Delta Q_q$ and $\Delta Q_{\bar q}$. 

Interestingly, we find that these inflections, and contrary to those of other quantities such as $M$, $\ell$, $\bar\ell$, and $n_q$ allow one to give a pretty global description of the phase diagram, including a connection between the singularities along the $\mu$-axis and the CEP, see Fig.~\ref{fig: phase dia from DQ}. More precisely, starting at the Silver-Blaze singularity $\smash{\mu=M_0}$, the inflections of $\Delta Q_q$ and $\Delta Q_{\bar q}$ both allow one to reach the CEP in the bulk of the phase diagram. Moreover, if we continue following these inflections from the CEP, we find that they follow quite closely the chiral crossover transition. In the case of $\Delta Q_q$, this is true down to rather small values of $\mu$, while for $\Delta Q_{\bar q}$ this is true until $\mu\simeq M_0/3$, after which the inflections deviate from the chiral ones and connect to the singularity at $\smash{\mu=-M_0/3}$. The situation is inverted for negative chemical potentials, as visualized in Fig.~\ref{fig: phase dia from DQ}.

We finally mention that the net quark number responses grow large in the vicinity and even diverge at the CEP. The divergence corresponds to the fact that $\partial_\mu\ell$ diverges while $\ell$ and $T$ remain finite. One possible physical interpretation is that, with a diverging correlation length, any multi-particle state can participate in screening the quark probe.

\section{Thermodynamic observables\label{sec: thermodynamics}}
One of the motivations for low-energy QCD models such as the PNJL model is that they provide a relatively simple framework for calculating quantities relevant for phenomenological applications, on top of just helping to understand the phase structure. The full equation of state, for instance, is especially important regarding certain astrophysical observations. Another example is the search for a critical endpoint through event-by-event fluctuations in heavy-ion collisions. In this case, ratios of various baryon number cumulants are also of interest.

\subsection{Basic formulas}
Let us give a quick overview of how to obtain the various relevant thermodynamic observables. The Landau free-energy (or grand potential) of a system of volume ${\cal V}$ at finite temperature $T$ and chemical potential $\mu$ is defined as
\beq
\Omega\equiv U-TS-\mu Q\,.
\eeq
It is such that
\beq
d\Omega=-p d{\cal V}-SdT-Qd\mu\,.
\eeq
In particular
\beq
p=-\left.\frac{\partial\Omega}{\partial{\cal V}}\right|_{T,\mu}\,.
\eeq
Then, if the system is homogeneous, one has $\smash{p=-\omega}$ with $\smash{\omega\equiv\Omega/{\cal V}}$ the Landau free-energy density. The latter obeys the simpler relations
\beq
\omega=\epsilon-Ts-\mu n_q\,,
\eeq
and
\beq
d\omega=-sdT-n_q d\mu\,,
\eeq
where $s$, $n_q$, and $\epsilon$ denote, respectively, the entropy, net quark number, and energy densities, related to $\omega$ as
\beq
s=-\frac{\partial\omega}{\partial T}\,, \quad n_q=-\frac{\partial\omega}{\partial \mu}\,,\label{eq: s nq}
\eeq
and
\beq
\epsilon=\left(1-T\frac{\partial}{\partial T}-\mu\frac{\partial}{\partial\mu}\right)\omega\,.
\eeq
Thus, to derive all relevant thermodynamic observables, it is enough to have a grasp on $\omega$. The latter can actually be obtained from the extremum of the potential that is
\beq
\omega=V(\sigma^{\rm ext},r_3^{\rm ext},r_8^{\rm ext})\,,
\eeq
with
\beq
0=\left.\frac{\partial V}{\partial r_3}\right|_{\sigma^{\rm ext},r_3^{\rm ext},r_8^{\rm ext}}\!\!\!\!\!=\left.\frac{\partial V}{\partial r_8}\right|_{\sigma^{\rm ext},r_3^{\rm ext},r_8^{\rm ext}}\!\!\!\!\!=\left.\frac{\partial V}{\partial \sigma}\right|_{\sigma^{\rm ext},r_3^{\rm ext},r_8^{\rm ext}}\,.\label{eq:extr}\nonumber\\
\eeq
We mention that due to the extremization conditions \eqref{eq:extr}, to take the $T$- and $\mu$-derivatives in \eqref{eq: s nq}, one only needs to focus on the explicit $T$- and $\mu$-dependencies while disregarding the implicit dependencies contained in $r_3^{\rm ext}$, $r_8^{\rm ext}$ and $\sigma^{\rm ext}$.

\subsection{\texorpdfstring{$p$, $s$ and $\epsilon$}{Pressure, entropy density and energy density}}
From the renormalization-group running of the QCD parameters, it is expected that, at very high temperatures, a QCD medium is essentially a gas of free, massless quarks and gluons. In this case, the pressure can be determined exactly, see e.g. \cite{Laine:2016hma}, and is given by the expression
\beq
    p_{SB} &=& \frac{\pi^2T^4}{90}\left(2(N_c^2-1)+\frac{7}{2}N_cN_f\right) \nonumber \\ 
    &&+\frac{N_cN_f}{6}\mu^2T^2+\frac{N_cN_f}{12\pi^2}\mu^4,\label{eq: SB}
\eeq
known as the Stefan-Boltzmann (SB) limit and from which the corresponding entropy density $s_{SB}$, energy density $\epsilon_{SB}$, and net quark number density $n_{q, SB}$ are readily derived.

\begin{figure}[t]
    \centering
    \includegraphics[width=0.9\linewidth]{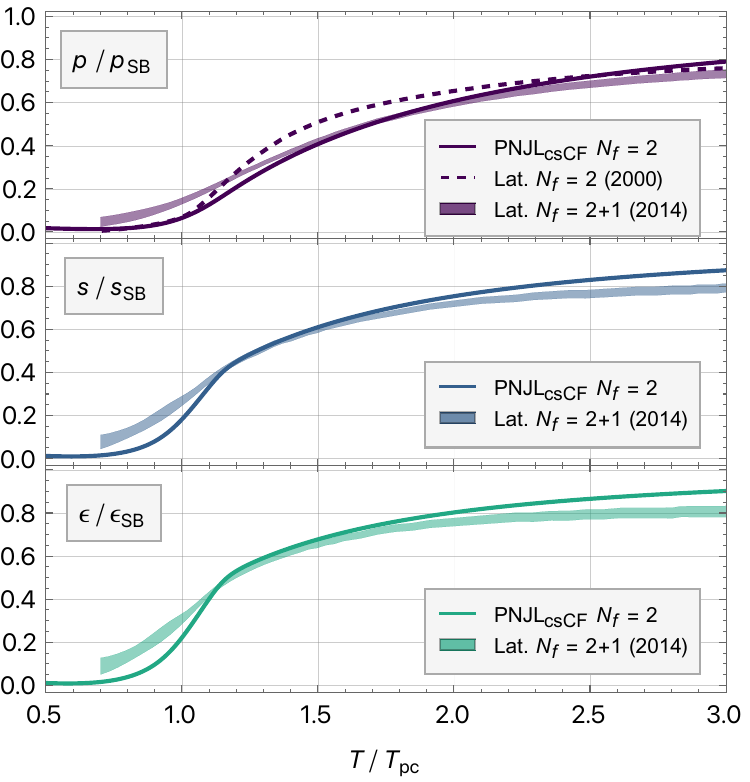}
    \caption{Pressure (top), entropy density (middle) and energy density (bottom) at ${\mu=0}$ normalized to their SB limits as functions of $T/T_{\rm pc}$ compared to continuum extrapolated ${N_f=2+1}$ lattice data \cite{Borsanyi:2013EoS} (bands), and ${N_f=2}$ pressure data on a ${16^3\times4}$ lattice \cite{Karsch:2000press} (dashed).}
    \label{fig: eos p s e}
\end{figure}

Fig.~\ref{fig: eos p s e} shows our results for the pressure $p/p_{SB}$, entropy density $s/s_{SB}$, and energy density $\epsilon/\epsilon_{SB}$ at ${\mu=0}$, rescaled by their respective SB limits as a function of $T/T_{\rm pc}$, compared to the continuum extrapolated lattice data with $\smash{N_f=2+1}$ of Ref.~\cite{Borsanyi:2013EoS} (rescaled by the three-flavor SB limit). We also compare the pressure to $\smash{N_f=2}$ lattice results from a single ${16^3\times4}$ lattice size, already given in terms of $T/T_{\rm pc}$, and with a temperature-scaling mass, see Ref.~\cite{Karsch:2000press}. While different pseudocritical temperatures may be derived for each quantity, for simplicity we used the common temperatures ${T_{\rm pc,\, PNJL}=175\,}$MeV for our results, and $T_{\rm pc,\, lat}=157\,$MeV for the lattice data. In other words, we do not fine-tune the temperature for better quantitative agreement, but rather compare the qualitative features, equal for other (reasonable) choices. We keep the same choices for the other figures in this section.

While the general shapes match very well for all quantities, the transition sets in more abruptly in the PNJL model. This is likely due to the missing hadronic degrees of freedom; recall that, according to the hadron resonance gas model, these degrees of freedom are needed to reproduce the lattice data below $T_{\rm pc}$. At high temperatures, the PNJL model converges faster to the SB limit than the lattice results.\footnote{We note that since the strange quark is heavier, the convergence in ${N_f=2+1}$ is expected to be slower, but this is not the only reason for the discrepancy.} However, there is still a clear gap between our results and the SB limit, even at high temperatures ${T\gtrsim3T_{\rm pc}}$. In the current model, this can actually be traced back mostly to the gauge contribution, since the quark contribution is already almost massless, and as such, closer to the SB limit.\footnote{The rapid growth of the Polyakov loops compared to the lattice results discussed in the previous section also makes the quark contribution grow faster.} In contrast, in the center-symmetric CF model, the finite gluon mass slows the convergence to the SB limit. We remind that our present implementation uses only one parameter in the glue sector, the CF mass, whereas most approaches utilize multi-parameter Ans\"atze.  The CF mass is determined by fitting the gluon propagator at zero temperature, in which case the center-symmetric Landau gauge boils down to the standard Landau gauge. One possibility would be to fit the gluon mass at each temperature, following Ref.~\cite{Kang:2022jbg}.

In QCD, the high-temperature behavior is described by the Hard-Thermal-Loop (HTL) resummation, giving a temperature-dependent mass $\smash{\sim g(T)T}$, see e.g. Refs.~\cite{Ghiglieri:2020dpq, Haque:2014rua, Laine:2016hma}, and thus slowing the convergence to the massless SB limit. This effect is not included in the current potential, but in practice, the CF mass term also leads to a reduced pressure at intermediate temperatures. An advantage of the (center-symmetric) CF model over other phenomenological Polyakov loop models is that one can include HTL resummation in the potential, which derives directly from the gluodynamics of the CF model, thereby improving the high-temperature behavior from first principles. This goes, however, beyond the scope of the current work, and we leave it for a future investigation.

We finally note that, like various other models \cite{Sasaki:2012bi, Reinosa:2014ooa, Canfora:2015yia, Kondo:2015noa, Folkestad:2018psc,Kroff:2018ncl}, the center-symmetric CF model does not correctly describe the $\smash{T\to 0}$ asymptotics of $p$, $s$, $\epsilon$ since these quantities do not show an exponential suppression with the temperature, see Ref.~\cite{Reinosa:2015gxn} for a thorough discussion. It remains an open question, beyond the framework of the CF model, to understand how to reconcile this expected exponential suppression with the observation, made on the lattice, that the Landau gauge gluon propagator is screened at small momenta, whereas the ghost propagator remains massless.

\subsection{Interaction measure and speed of sound}

From combinations of the basic functions $p$, $s$ and $\epsilon$, one can obtain other interesting functions that tell us about refined properties of the system.

\begin{figure}[t]
    \centering
    \includegraphics[width=0.9\linewidth]{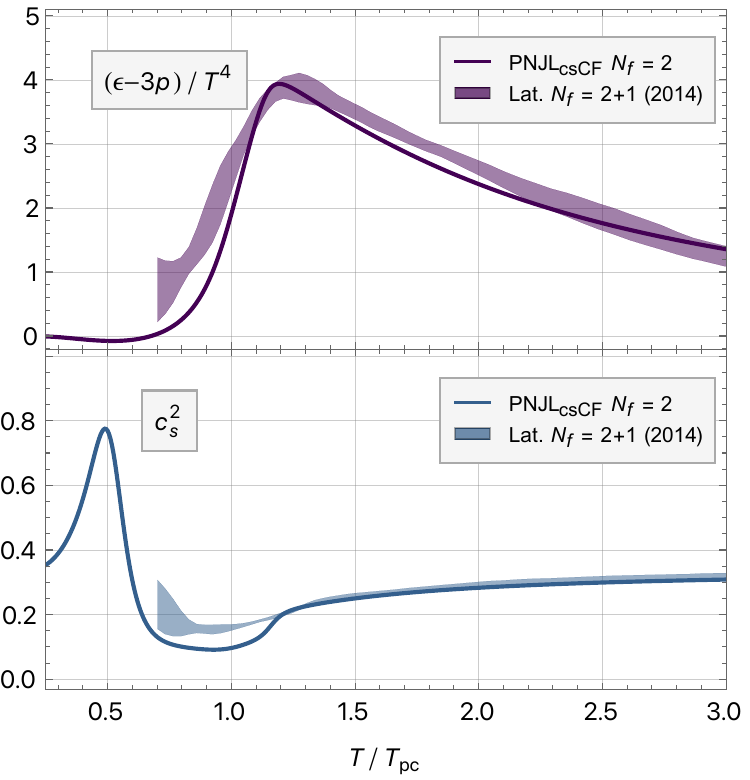}
    \caption{Temperature normalized trace anomaly ${(\epsilon-3p)/T^4}$ (top), and speed of sound squared (bottom) at ${\mu=0}$ as functions of $T/T_{\rm pc}$ compared to continuum extrapolated ${N_f=2+1}$ lattice data \cite{Borsanyi:2013EoS} (bands).}
    \label{fig: eos I cs}
\end{figure}

The difference $\smash{\epsilon-3p}$ corresponds to the trace of the energy-momentum tensor. It vanishes in the Stefan-Boltzmann limit, as it should for a scaleless, non-interacting system such as the free quark-gluon gas described by the Stefan-Boltzmann pressure. In QCD, due to the strong quark-gluon interactions, a scale violation is generated. The deviation of QCD from a conformal theory can then be measured by $\epsilon-3p$, commonly referred to as the trace anomaly or interaction measure.

The speed of sound $c_x$ in the medium relates to the pressure change due to a change in energy density
\begin{equation}
    c_x^2=\left.\frac{\partial p}{\partial\epsilon}\right|_x,
\end{equation}
where, depending on the fixed quantity $x$ at which the derivative is calculated, different ``sound speeds" may be considered. The ideal propagation of fluctuations in a medium is adiabatic, so one usually keeps the entropy density fixed, thus leading to $c_s$. For the near-perfect fluid created in heavy-ion collisions, total energy and net baryon number conservation mean that the meaningful definition should use the conserved entropy density per baryon instead, leading to $c_{s/n_B}$. At small $\mu$ the entropy per baryon number contours are almost parallel to the temperature axis, and in the limit $\mu\to0$ the speed of sound becomes $c_s=c_{\mu=0}$, and we may simply approximate
\begin{equation}
    c_{s/n_B}^2\simeq c_\mu^2=\frac{\partial p}{\partial\epsilon}\bigg|_\mu=\frac{\frac{\partial p}{\partial T}\big|_\mu}{\frac{\partial\epsilon}{\partial T}\big|_\mu}.
\end{equation}
This is also where lattice simulations are viable so we can compare the results in that limit. However, heavy-ion collisions take place at finite $\mu$, and for increasing $\mu$, the speeds differ, and more care needs to be taken, see e.g. Ref.~\cite{He:2022Soundspeed}. 

In Fig.~\ref{fig: eos I cs} we compare the normalized trace anomaly ${(\epsilon-3p)/T^4}$ and squared speed of sound $c_s^2$ at ${\mu=0}$ to continuum extrapolated ${N_f=2+1}$ lattice data \cite{Borsanyi:2013EoS}, as functions of $T/T_{\rm pc}$. The shape and peak height of the trace anomaly are reproduced very well, with the behavior at high temperatures starting to differ slightly, as was also the case for the other quantities. There is a very small temperature region where it becomes negative, which may also be related to the non-monotonicity of the pressure at low temperatures in the CF model. The shape of the sound speed is also well reproduced, and for ${T\gtrsim1.2T_{\rm pc}}$ the quantitative agreement is also remarkable, with clear convergence towards the expected conformal value ${c_s^2=1/3}$.

\begin{figure}[t]
    \centering
    \includegraphics[width=0.9\linewidth]{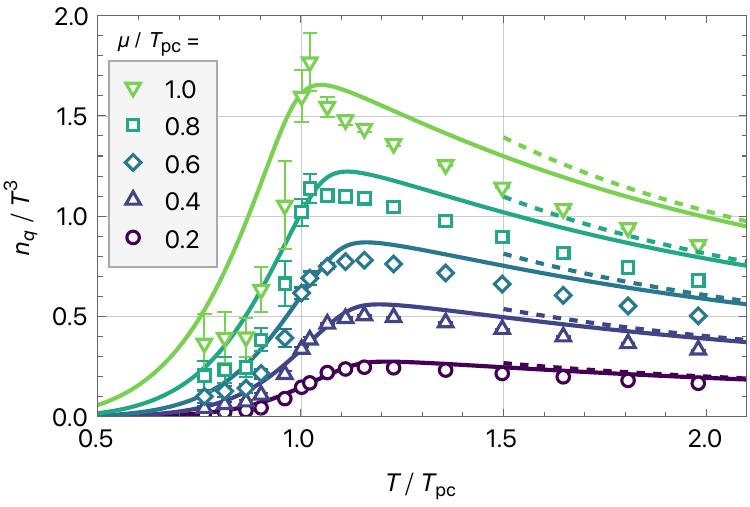}
    \caption{Quark number density $n_q/T^3$ as a function of $T/T_{pc}$ for five $\mu$'s up to ${T_{pc}}$ (solid lines), compared to ${N_f=2}$ lattice data \cite{Allton:2003vx} (symbols). The dashed lines show the SB limit, which is approached faster in the PNJL model.}
    \label{fig: nq}
\end{figure}

\subsection{Net quark number density}
Fig.~\ref{fig: nq} displays our results for the normalized net quark number density $n_q/T^3$ at various, not too large, chemical potentials, compared to ${N_f=2}$ lattice data from a Taylor expansion \cite{Allton:2003vx}. While the general shape is well reproduced, the PNJL model results systematically overestimate $n_q/T^3$ above $T_{\rm pc}$, by around 15\% at ${T=2T_{\rm pc}}$. This can directly be traced back to the behavior of the first Taylor expansion coefficient, which is a similar amount away from the SB limit in Ref.~\cite{Allton:2003vx}, and plateaus around this distance. Instead, in the PNJL model, $n_q$ is already much closer to the SB limit, which is marked by the dashed lines in Fig.~\ref{fig: nq}. This is related to our discussion of the other thermodynamical quantities' overly fast convergence to the SB limit, particularly for the quark contributions, which have the strongest impact in ${n_q=-\partial_\mu\omega}$, as the only $\mu$-dependence in $\omega$ comes from the quark contributions.

In Sec.~\ref{subsec: M(T to 0)} we discussed the onset of a dense phase at ${\mu=M_0}$, even before the first-order chiral transition happens. The net quark number density $n_q$ remains finite even as $ T\to 0$. In fact, from the $T=0$ potential \eqref{eq: M potential T to 0}, we can get the exact form of the net quark number density for $|\mu|>M$ at ${T=0}$
\begin{equation}\label{eq: nq(T=0)}
    n_q^{T=0}(\mu)=\frac{N_f}{\pi^2}\left(\mu^2-M(\mu)^2\right)^{3/2},
\end{equation}
which is the well-known density of a massive fermion system expressed using the Fermi momentum $\propto\sqrt{\mu^2-M^2}$. Note that expanding this for small masses, which is reasonable after the first-order transition, gives essentially the zero $T$ limit of the net quark number density in the free quark gas (SB) with mass-dependent corrections
\begin{equation}
n_q^{T=0}=n_{q,\, SB}^{T=0}\left(1-\frac{3}{2}\frac{M^2}{\mu^2}+\mathcal{O}(M^4/\mu^4)\right).
\end{equation}
It is convenient to plot the net quark number density normalized by the SB result $n_{q,\, SB}=N_f(\mu T^2+\mu^3/\pi^2)$, since it then stays between 0 and 1 throughout the phase diagram.

\begin{figure}[t]
    \centering
    \includegraphics[width=0.9\linewidth]{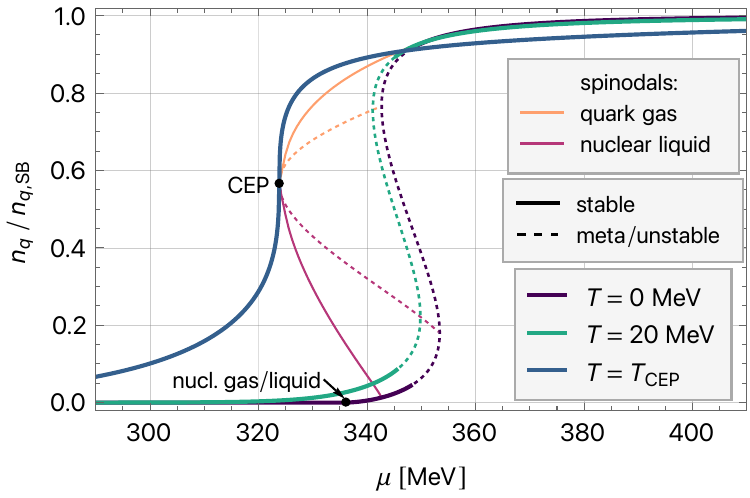}
    \caption{Quark number density normalized to its SB value $n_q/n_{q, SB}$ as a function of $\mu$ for three temperatures at the spinodal region, where the dashed lines denote the density of the metastable or unstable phase solutions. The thin lines originating from the CEP and ending at the $T=0$ values mark the densities of the chirally broken($\sim$nuclear liquid, magenta) and chirally restored($\sim$quark gas, orange) phases at the edges of the spinodal/coexistence region.}
    \label{fig: nq-norm}
\end{figure}

In Fig.~\ref{fig: nq-norm} we show $n_q/n_{q,\, SB}$ as a function of $\mu$ for the temperatures delimiting the spinodal region and one temperature inside it. The ${T=0}$ limit was calculated using the relation \eqref{eq: nq(T=0)}, and gives a nice cross-check for our numerical results at low $T$, which indeed converge to this limit. The thin and dashed lines connecting the CEP to the ${T=0}$ curve show the normalized density along the spinodals as the temperature decreases. At the left (lower $\mu$) spinodal, the constituent quark solution is stable, which should correspond to the nuclear liquid, and at the right one, the (approximately) chirally restored quarks are stable. Of course, the present model does not account for diquarks, so here this phase is just a gas of quarks and gluons. The PNJL model also does not account for nucleons, and thus the transition at $\mu=M_0$ is at best a very crude approximation of the nuclear liquid gas transition, in terms of the onset of a finite-density phase of deconfined but chirally broken quarks.

\subsection{Susceptibilities}
\begin{figure}[t]
    \centering
    \includegraphics[width=0.9\linewidth]{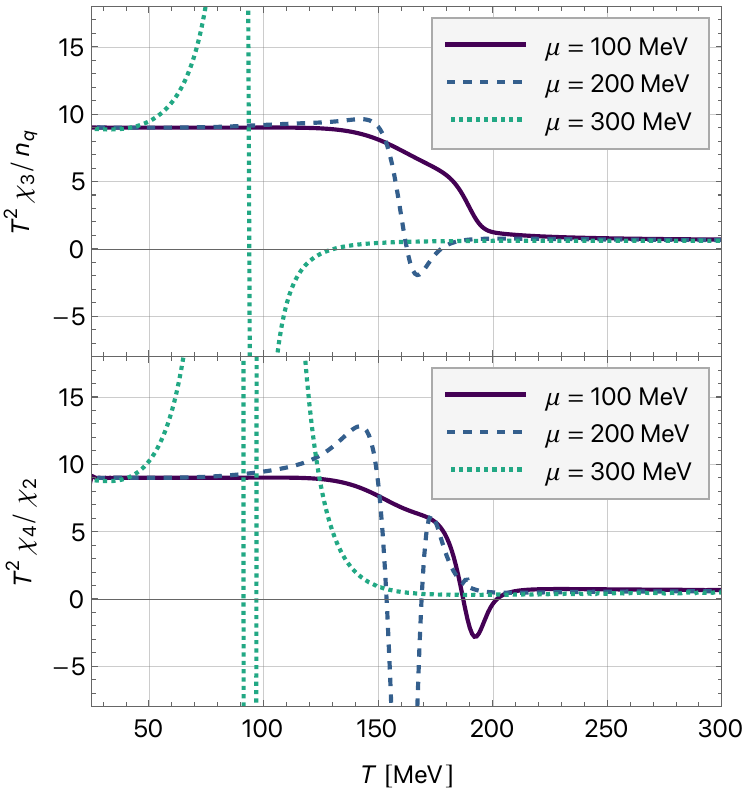}
    \caption{Ratios of quark number susceptibilities separated by two $\mu$ derivatives as a function of the temperature for various chemical potentials.}
    \label{fig: Chi-ratios}
\end{figure}

Various proton number cumulants have been measured through event-by-event fluctuations in heavy ion collisions \cite{STAR:2025zdq}, and serve as a proxy for baryon number cumulants, which are sensitive to a critical point \cite{Bzdak:2016sxg}. To make predictions for the deviation from a non-critical baseline, one needs to convert the simple $T$- and $\mu$-dependence of the baryon number cumulants to match the experimental setup, where larger $\mu$ correspond to decreasing collision energies. Here we settle for showing the temperature dependence of the ratios $\chi_3/n_q$ and $\chi_4/\chi_2$ multiplied by $T^2$ to account for the dimensionality, at various chemical potentials, see Fig.~\ref{fig: Chi-ratios}. We defined ${\chi_n=(\partial_\mu)^n p}$ and ${n_q=\chi_1}$, where we note that various conventions for the naming of these quantities exist in the literature.

In the confined phase, the ratios $\smash{T^2\chi_{n+2}/\chi_n}$ converge to $\smash{9=3^2}$ as every $\mu$-derivative only brings factors of $\smash{N_c=3}$. Instead, at high temperatures, the (two shown, lowest order) ratios converge towards $6/\pi^2$, which corresponds to the SB limit. Around the crossover, they vary smoothly between these values, but as $\mu$ is increased and the critical endpoint approached, they start being singular and changing signs, which should have a measurable impact in heavy-ion collisions if the fireball's evolution through the phase diagram passes close enough to the CEP. As visible in Fig.~\ref{fig: Chi-ratios}, this effect becomes stronger for the higher-order ratios, which peak more strongly and change sign more often, and already at lower $\mu$.

\section{Conclusions}\label{sec: conclusions}
We have proposed to study certain aspects of the QCD phase diagram from a phenomenological perspective, merging two popular models, the Nambu--Jona-Lasinio model on the one hand, which captures the dynamics behind spontaneous chiral symmetry breaking, and the center-symmetric Curci-Ferrari model on the other hand, which captures the dynamics behind the confinement/deconfinement transition. This approach, which belongs to the large class of Polyakov--Nambu--Jona-Lasinio type models, stands out through the fact that the modeling of the glue sector involves only one phenomenological parameter and can be treated within a weakly coupled expansion.

Generally, we obtain a good qualitative description of the main expected features of the QCD phase diagram, including the presence of a critical endpoint and the behavior of thermodynamic quantities. Some of our results agree quantitatively with other approaches, such as lattice or functional methods, but clearly not all. We also find that our approach is basically equivalent to other PNJL models with different Ans\"atze for the Polyakov loop potential, but with a reduced number of phenomenological parameters to adjust. When comparing these various models, we find differences in the behavior of the Polyakov loops at low temperatures and high densities $\smash{|\mu|>M_0}$, but we argue that this different behavior should not affect the physical interpretation.

To this purpose, we supplement the analysis of the Polyakov loops with related quantities, in particular the net quark number responses of the medium to the addition of an external quark and anti-quark probe. Not only do these quantities allow one to disambiguate the interpretation of the region $\smash{|\mu|>M_0}$, but they reveal intriguing behaviors in the region $\smash{|\mu|<M_0}$, which generalize those already obtained in the case of heavy-quark QCD \cite{MariSurkau:2025MesBar}. They are reminiscent of the fact that, in the confined phase of the system, the external probe is absorbed into states with the same net quark number as mesons or baryons. We also argue that these quantities are useful beyond the zero-temperature limit, and allow one to probe the QCD phase diagram, while revealing interesting connections between the CEP in the bulk of the phase diagram and singularities along the $\mu$-axis.

Our analysis could be improved in various directions. First, including a heavier strange quark, $\smash{N_f=2+1}$, should not be too difficult. In particular, at zero temperature, a similar analysis can be performed separately for the light and strange contributions. We expect there to be a singularity at $M_{l,0}$, along with the first-order jump related to the light quarks, but which of course affects all order parameters. Then, a singularity related to $M_{s,0}$, and after that, common parameterizations seem to favor just a crossover for the strange quark mass, rather than another first-order jump \cite{Fukushima:2008wg}. We note that some care should be taken in how the strangeness chemical potential is treated, and whether one enforces strangeness neutrality, as relevant for heavy-ion collisions. 

At large chemical potentials, the analysis should be extended to include color-superconducting degrees of freedom \cite{Son:1998uk, Roessner:2006xn}, as they are known to dominate at low temperatures, see e.g.~\cite{Braun:2019aow}. We expect similar results in terms of the interpretation of confined phases, but changes to the NJL part due to the interplay of the chiral and diquark condensates. In particular, it will be interesting to analyze the behavior of the net quark number responses in phases dominated by di-quarks. This is left for future work, however. Of course, a combined treatment in $\smash{N_f=2+1}$ with the inclusion of diquarks would be interesting for the sake of a more realistic QCD analysis.

The above points are all related to the shortcomings of the simple NJL model we used for the quark sector. Other possible improvements include refining the modeling of the gauge sector through the evaluation of two-loop expansion and/or HTL corrections to the glue potential, or replacing the NJL model by a more first-principle approach based on the rainbow-improved expansion scheme \cite{Pelaez:2017bhh, Pelaez:2020ups}. Both types of improvements can be implemented within the Curci-Ferrari framework, and work is already in progress along these lines.

\acknowledgements{We would like to thank G. Endrődi for useful comments concerning the renormalization of the Polyakov loop.}

\appendix

\section{Parameter dependence}\label{app: params}
Both the gauge (P) and quark (NJL) sectors of the PNJL model contain parameters. Throughout the main text, we kept a constant choice, but we now analyze the impact of varying the parameters of each sector alone and together.

\subsection{Glue sector}
In the CF model, the glue/Polyakov loop potential depends on the gluon mass $m$, the gauge coupling $g$, and the renormalization scale $\xi$ and scheme, with $m$ and $g$ implicitly depending on the scale and scheme. As the relevant temperature scales are quite small, we use a Landau-pole-free infrared safe scheme, but we note that other schemes exist and give equivalent results in the glue sector \cite{MariSurkau:2024zfb} and for heavy quarks \cite{MariSurkau:2025dfo}, and we have confirmed that this is also the case for the PNJL model. 

We then consider two types of parameter variations. One comes from changing the initial parameters $m(\xi_0)$ and $g(\xi_0)$ at some scale $\xi_0$, and the other from changing the renormalization scale $\xi$, and consequently the running parameters $m(\xi)$ and $g(\xi)$. 

For the first type of variation, there are, in principle, various ways one can obtain the parameters. The main text used ones obtained by fitting one-loop expressions for the YM propagators to corresponding lattice results, giving $m_g(\xi_0)\simeq390\,$MeV and $g(\xi_0)\simeq3.7$ at $\xi_0=1\,$GeV \cite{Tissier:2011ey}. It is reasonable to argue that when considering light quarks, as done here, the fits should instead be to propagators that include the effect of such quarks. Spontaneous chiral symmetry breaking has been studied in the vacuum with the CF model using a double expansion in $g$ and $N_c^{-1}$. The resulting simultaneous fits to gluon and quark propagators yield values of around $m_g(\xi_0)\simeq210\,$MeV and $g(\xi_0)\simeq2.45$ at $\xi_0=1\,$GeV \cite{Pelaez:2020ups}. We note that the quality of the fit is slightly worse here than in pure YM, because the lattice quark/gluon propagators favor slightly lower/higher gluon masses in a one-loop fit, leading to a compromise.

In Table~\ref{tab: T glue fits} we summarize the pseudocritical temperatures at vanishing chemical potential and the locations of the CEP that are obtained with the two different initial parameter choices (without running the scale). We find that the temperatures decrease, in agreement with the expectation from lattice simulations that the YM transition happens at a higher temperature than that of QCD.

\begin{table}[h]
    \centering
    \begin{tabular}{|c||c|c|c|}\hline
           {\color{gray}[MeV]} & $T^\ell_{pc}(\mu=0)$ & $T^\sigma_{pc}(\mu=0)$ & $(\mu_{CEP},T_{CEP})$ \\ \hline\hline
        YM fit & 174.6 & 197.4 & (323.8,\,64.7) \\ \hline
        QCD fit & 136.0 & 184.7 & (325.8,\,56.8) \\\hline
    \end{tabular}
    \caption{Pseudocritical temperatures at ${\mu=0}$ and location of the CEP for the two different parameter choices in the gauge sector.}
    \label{tab: T glue fits}
\end{table}

For the second type of variation, we utilize the running of parameters as determined from the perturbative pure gauge sector, see e.g. \cite{Tissier:2011ey, MariSurkau:2024zfb}, meaning the effect of the quarks is not incorporated into the running. This is an unavoidable shortcoming of the PNJL model. Since the quark sector is not renormalizable, the RG running can not be performed consistently. That we can even consider running can be seen as a strength of the CF model compared to other models of the gauge sector, which are purely phenomenological and not obtained from a renormalizable Lagrangian. 

Concerning the pseudocritical temperatures at ${\mu=0}$, varying the scale across a very large range of values $\sim$0.5-5\,GeV produces changes of under 5\,MeV for $T_{pc}^\ell$ and under 1\,MeV for $T_{pc}^\sigma$ for the YM parameter set, and even less for the QCD set. This speaks for the validity of the perturbative approximation in the CF model, as the observables should not depend on the arbitrary renormalization scale. More significant variations appear in the regimes of small $\xi$, with the deconfinement or chiral crossover temperatures decreasing by up to tens of or ten Mev. This is somewhat expected, as in this regime the coupling and mass vary the strongest and have a peak, making the perturbative expansion less accurate in this case. Going to scales below the gluon mass is questionable anyway, as the dynamics should freeze out there.

\begin{figure}[t]
    \centering
    \includegraphics[width=.95\linewidth]{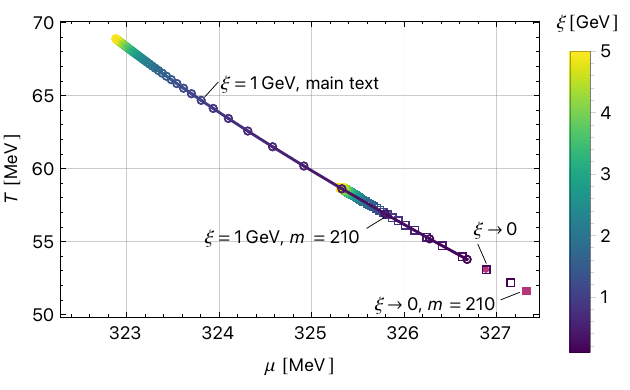}
    \caption{Location of the CEP as we vary the renormalization scale ${\xi\in[0.1,5]\,}$GeV used for the IR-safe glue potential with the two initial parameter choices. The circles connected by a line are the main text (pure YM) parameters, and the squares are the lower gluon mass ${m=210\,}$MeV parameters from a QCD fit. The symbols correspond to steps of 0.1\,GeV, and the filled circle and square show the corresponding ${\xi=0}$ limits.}
    \label{fig: CEP of s}
\end{figure}

To illustrate the renormalization scale dependence, we show, in Fig.~\ref{fig: CEP of s}, how the CEP varies with the scale $\xi$ for both of the initial parameter choices above. The scale is varied in the interval $[0.1,5]\,$GeV, and we use symbols to mark regular intervals of 100\,MeV because the displacement of the CEP is non-linear in $\xi$. We see again that the position varies the fastest around where the coupling peaks (near ${\xi\sim0.3\text{-}0.4\,}$GeV), converges in the IR, and seems to converge in the UV as well. We note that the displacement of the CEP seems to be ``universal", falling on the same line independent of the considered RG trajectories, just with different IR and UV limits.

\subsection{NJL sector}
Throughout this paper we used parameters that give $\smash{m_\pi\simeq138.2\,}$MeV, $\smash{f_\pi\simeq93.1\,}$MeV, $\smash{\langle\bar\psi_l\psi_l\rangle^{1/3}\simeq247\,}$MeV, and $\smash{M\simeq335.9\,}$MeV. These are fairly close to the PDG values ${m_{\pi^{\pm}}=139.57\,}$MeV, ${m_{\pi^0}=134.98\,}$MeV, and ${f_{\pi}=92.07(85)\,}$MeV \cite{ParticleDataGroup:2024cfk, Aoki2016FLAGreview}, but considering alternative parameter choices may improve the agreement, and should also be done to test the sensitivity of the results to the model.

Since we work in the isospin-symmetric case and neglect QED interactions, for consistency, we could use $m_\pi$ and $f_\pi$ found in this limit, which, of course, do not correspond exactly to the experimental values measured away from this limit. However, lattice QCD simulations typically operate in this limit. Hence, numerous results for these quantities exist in the literature, which closely align with the physical findings, confirming that this approximation is valid. Technically, using these also makes our comparisons to lattice QCD simulations more consistent.\footnote{Older (two-flavor) lattice simulations often used unphysical pion masses, so another way of cross-testing the model would be to fit the parameters anew according to the pion mass used for simulations.} 
QED self-energy effects lead to an increase in the charged pion masses while the isospin breaking effects slightly decrease them, so for isospin-symmetric QCD, the flavor lattice averaging group (FLAG) \cite{Aoki2016FLAGreview} estimates the pion masses ${m_\pi=134.8(3)\,}$MeV. They also provide values for the ${N_f=2}$ light quark condensate ${\langle\bar\psi_l\psi_l\rangle^{1/3}=266(10)\,}$MeV \cite{Cichy:2013gja, Brandt:2013dua, ETM:2009ztk, Engel:2014eea} (this value increases by ${\sim6\,}$MeV for ${N_f=2+1}$). The parameter set ${m_l=4\,}$MeV, ${\Lambda=768.5\,}$MeV, ${G=6.40\,\text{GeV}^{-2}}$ reproduces these values well. We call it NJL II, and the above choice NJL I.

\begin{figure}[t]
    \centering
    \includegraphics[width=0.9\linewidth]{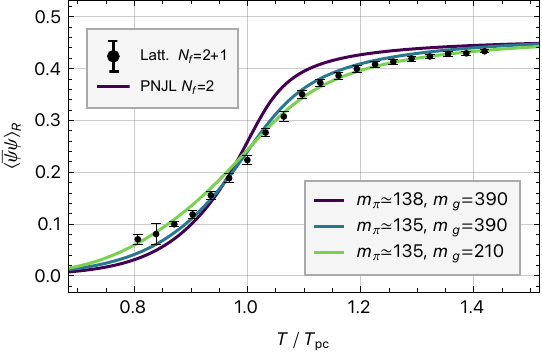}
    \caption{Renormalized chiral condensate (${N_f=2}$) found with different parameter choices (in MeV) in the NJL and gauge sector, compared to the lattice data from \cite{Borsanyi:2010ordPar} (${N_f=2+1}$). The dark line corresponds to the choice of the main text, the others to varying just the NJL or both parameters.}
    \label{fig: param depend chiral cond}
\end{figure}

\begin{table}[h]
    \centering
    \begin{tabular}{|c||c|c|c|}\hline
           {\color{gray}[MeV]} & $T^\ell_{pc}(\mu=0)$ & $T^\sigma_{pc}(\mu=0)$ & $(\mu_{CEP},T_{CEP})$ \\ \hline\hline
        NJL I, YM & 174.6 & 197.4 & (323.8,\,64.7) \\ \hline
        NJL II, YM & 164.3 & 183.8 & -- \\\hline
        NJL II, QCD & 127.9 & 168.8 & -- \\\hline
    \end{tabular}
    \caption{Pseudocritical temperatures at ${\mu=0}$ and location/existence of the CEP for the two different parameter choices in the NJL sector with the YM gluon mass and in the last line with the QCD gluon mass.}
    \label{tab: T njl params}
\end{table}

In Table~\ref{tab: T njl params} we summarize the pseudocritical temperatures at vanishing chemical potentials obtained with the two choices and the glue parameters of the main text (YM), and in the last line we also show the results when using both alternative parameter sets. The NJL II set pushes the transition down, closer to the lattice results that use the same pion mass, but still not in agreement. Using both the NJL II set and the QCD fitted gauge parameters gives a chiral transition temperature that is only about 10 MeV from the lattice estimates, but the deconfinement transition seems to be pushed to rather low temperatures. In Fig.~\ref{fig: param depend chiral cond} we compare the effect of different combinations of NJL and gauge parameters on the renormalized chiral condensate (see Eq.~\eqref{eq: renorm chir cond}) at $\mu=0$. Indeed the NJL II set follows the lattice results \cite{Borsanyi:2010ordPar} practically within error bars, and the lower gluon mass (QCD fit) seems to smoothen the transition, giving the best agreement.

With the NJL II parameter set we find however another unexpected result. The CEP disappears completely from the phase diagram, and the transition is a crossover even along the $\mu$-axis at ${T=0}$. It seems that the existence of a CEP in the (P)NJL model is rather sensitive to the parameter values.\footnote{The existence is fully determined by the NJL parameters, as those decide whether there is a crossover or a first-order transition at ${T=0}$. Since (for ${m_l>0}$ and small enough $G$
) there is a crossover at ${\mu=0}$ in the latter case a CEP must exist in the phase diagram. Its position is then also influenced by the gauge parameters.} If we vary the current quark mass while enforcing the same pion mass and decay constant, i.e. adjusting $\Lambda$ and $G$, we find that for ${m_\pi\simeq135\,}$MeV a CEP will only appear for quark masses above ${m_l\simeq4.9\,}$MeV. For ${m_\pi\simeq138\,}$MeV (as in the main text) the CEP disappears below ${m_l\simeq5.1\,}$MeV. Those values give light quark condensates roughly within the error bars of the lattice results. Actually, the NJL I set commonly used in the literature gives a condensate that is further away (too small) from the lattice results. We conclude that no clear statement about the position or even existence of a CEP in the QCD phase diagram can be made from the (P)NJL model, and more sophisticated calculations or models of the quark sector are necessary.

\section{Relation between order parameters}\label{app:CEP}
Consider two (possibly multidimensional) order parameters $x_i$ and $y_a$ whose value as a function of $T$ and $\mu$ is obtained by extremizing a potential $V(x_i,y_a)$. We imagine that this potential depends on some implicit parameter $\lambda$ and that, for certain values of this parameter, the potential is invariant under a transformation $\smash{x_i\to X_{ij} x_j}$ or a transformation $\smash{y_a\to Y_{ab} y_b}$ , or both, but that, away from these particular values, these symmetries are explicitly broken. We assume for simplicity that $X-{\rm Id}$ and $Y-{\rm Id}$ are invertible. In this way, when the symmetry $X$ (resp. $Y$) is manifest and the extremum is invariant, $\smash{X_{ij}x_j=x_i}$ (resp. $\smash{y_a=Y_{ab}y_b}$), we have necessarily $\smash{x_i=0}$ (resp. $\smash{y_a=0}$).

We now study the appearance of singularities in the values of the order parameters $x_i$ and $y_a$ as $T$ and/or $\mu$ are varied, such as first-order transition discontinuities, spinodals, or critical end-points. We would like to argue that, for generic values of the parameter $\lambda$, that is, when the symmetries are explicitly broken, the same singularities need to be associated with both order parameters, and that, only when a symmetry protects one of the two order parameters, independent singularities can occur.

Since first-order transitions occur between spinodals and because critical end points are usually located at the end of first-order transition lines, we shall focus here on spinodals. A similar analysis can be done to treat critical endpoints.

\subsection{Generic case}
Consider first the generic case where both symmetries introduced above are explicitly broken at the level of the potential. Let us now explain how we would associate a spinodal to each of the order parameters, before showing that these spinodals need to be the same.

For the order parameter $x_i$, we would first introduce its reduced potential as
\beq
V_x(x_i)=V(x_i,y_a(x_i))\,,
\eeq
where $y_a(x_i)$ is defined from
\beq
0=\frac{\partial V}{\partial y_a}\,.\label{eq:Vya}
\eeq
Now a spinodal point for $x_i$ corresponds to a solution in the $(\mu,T)$-plane of the equations
\beq
0=\frac{\partial V_x}{\partial x_i}\,, \quad 0={\rm det}\,\frac{\partial^2V_x}{\partial x_i\partial x_j}\,,\label{eq:spinodalx}
\eeq
which need to be solved together with (\ref{eq:Vya}). Note that we have as many equations as they are components for $x_i$ and $y_a$, plus one extra equation. This corresponds, at most to a curve in the $(\mu,T)$-plane. A similar procedure could be considered to define the spinodal points associated to $y_a$, from the corresponding reduced potential $V_y(y_a)$. Let us now argue that the so-defined spinodal points, associated respectively with $x_i$ and $y_a$, coincide in the generic case considered here.

We first note that, owing to Eq.~(\ref{eq:Vya}),
\beq
\frac{\partial V_x}{\partial x_i}=\frac{\partial V}{\partial x_i}+\frac{\partial V}{\partial y_a}\frac{\partial y_a}{\partial x_i}=\frac{\partial V}{\partial x_i}\,.\label{eq:same}
\eeq
From this it follows that the first condition in (\ref{eq:spinodalx}), together with (\ref{eq:Vya}) is equivalent to
\beq
0=\frac{\partial V}{\partial x_i}\,, \quad 0=\frac{\partial V}{\partial y_a}\,,
\eeq
which we can rewrite compactly as
\beq
0=\frac{\partial V}{\partial z_h}\,,\label{eq:Vz1}
\eeq
where $z_h=(x_i,y_a)$. 

Let us now analyze the second equation in (\ref{eq:spinodalx}) in the presence of the other equations defining the spinodal of $x_i$. Starting from (\ref{eq:same}), we obtain
\beq
\frac{\partial^2 V_x}{\partial x_i\partial x_j}=\frac{\partial^2 V}{\partial x_i\partial x_j}+\frac{\partial^2 V}{\partial x_i\partial y_a}\frac{\partial y_a}{\partial x_j}\,,\label{eq:B8}
\eeq
and, again from Eq.~(\ref{eq:Vya}), we find
\beq
0=\frac{\partial y_b}{\partial x_i}\frac{\partial^2 V}{\partial y_b\partial y_a}+\frac{\partial^2 V}{\partial y_a\partial x_i}\,.\label{eq:B9}
\eeq
Now, in the generic case, it is natural to assume that, along the spinodal line defined by Eqs.~(\ref{eq:Vya}) and (\ref{eq:spinodalx}), and except maybe at certain points, we have $\smash{{\rm det}\,\partial^2V/\partial y_a\partial y_b\neq 0}$. This is because, otherwise, we would have an additional condition that would determine $T$ and $\mu$. We can then assume that $\partial^2V/\partial y_a\partial y_b$ is invertible and write (\ref{eq:B9}) as
\beq
\frac{\partial y_a}{\partial x_j}=-\left(\frac{\partial^2 V}{\partial y_a\partial y_b}\right)^{-1}\frac{\partial^2 V}{\partial y_b\partial x_j}\,,
\eeq
which, when plugged back into (\ref{eq:B8}) leads to
\beq
\frac{\partial V_x}{\partial x_i\partial x_j} & = & \frac{\partial^2 V}{\partial x_i\partial x_j}-\frac{\partial^2 V}{\partial x_i\partial y_a}\left(\frac{\partial^2 V}{\partial y_a\partial y_b}\right)^{-1}\frac{\partial^2 V}{\partial y_b\partial x_j}\,,\nonumber\\
\eeq
so that the second equation in (\ref{eq:spinodalx}) rewrites
\beq
0={\rm det}\,\left[\frac{\partial^2 V}{\partial x_i\partial x_j}-\frac{\partial^2 V}{\partial x_i\partial y_a}\left(\frac{\partial^2 V}{\partial y_a\partial y_b}\right)^{-1}\frac{\partial^2 V}{\partial y_b\partial x_j}\right].\nonumber\\
\eeq
Again, because we can assume that $\partial^2V/\partial y_a\partial y_b$ is invertible, we do not lose any information if we multiply this equation by ${\rm det}\,\partial^2V/\partial y_a\partial y_b$. Then, using Schur's formula for the determinant of a block matrix, we recognize
\beq
0={\rm det}\,\frac{\partial^2 V}{\partial z_h\partial z_k}\,.\label{eq:Vz2}
\eeq
We have thus arrived at the following conclusion: the spinodal line associated with $x_i$, which was originally defined from its reduced potential $V_x(x_i)$, can be equivalently defined as a spinodal line for the potential $V(z_h)$, see Eqs.~(\ref{eq:Vz1}) and (\ref{eq:Vz2}), where any particularization of $x_i$ against $y_a$ is lost. This shows that the notion of spinodal, in the considered generic case, is the same for all order parameters, and, actually if we would had started, instead, from the definition of the spinodal for $y_a$ in terms of its reduced potential $V_y(y_a)$, we would have ended up with the same equations (\ref{eq:Vz1}) and (\ref{eq:Vz2}).

\subsection{Symmetric case}
In the particular case where the potential is symmetric under any of the symmetries $\smash{x_i\to X_{ij}x_j}$ or $\smash{y_a\to Y_{ab}y_b}$ described above, the above discussion does not apply, as we now discuss.

Suppose, for instance, that the potential is invariant under $\smash{x_i\to X_{ij}x_j}$. Then, in the symmetric phase, the relevant extremum is at $\smash{x_i=0}$. Moreover, at this point, the symmetry implies
\beq
(X_{ij}-\delta_{ij})\left.\frac{\partial^2V}{\partial x_j\partial y_a}\right|_{x=0,y}=0\,,
\eeq
and thus
\beq
\left.\frac{\partial^2V}{\partial x_j\partial y_a}\right|_{x=0,y}=0\,,\label{eq:decoupled}
\eeq
since we have assumed that $X-{\rm Id}$ was invertible. In this case the condition (\ref{eq:Vz2}) rewrites
\beq
0={\rm det}\,\frac{\partial^2 V}{\partial x_i\partial x_j}\times {\rm det}\,\frac{\partial^2 V}{\partial y_a\partial y_b}\,,
\eeq
which means that
\beq
0={\rm det}\,\frac{\partial^2 V}{\partial x_i\partial x_j}\,,
\eeq
or
\beq
0={\rm det}\,\frac{\partial^2 V}{\partial y_a\partial y_b}\,,
\eeq
and thus one of the assumptions we made in the previous section to show the equivalence of the spinodals does not apply. Equation (\ref{eq:decoupled}) means that for all questions that involve at most second derivatives in the symmetric phase of $x$, such as the determination of the spinodal line in the symmetric phase, the two order parameters are decoupled from each other.

\subsection{Equations for the spinodals and CEP}
For completeness, we give the equations to find the spinodal lines and the CEP in the current model, where there are three order parameters, $\sigma,\,r_3,\,r_8$, related to the chiral, center and charge conjugation symmetries. At the spinodals, the metastable and unstable extrema merge, so the determinant of the Hessian of the potential needs to vanish at this point
\beq
   0&=&\det \mathcal{H}(V(\sigma,r_3,r_8))
   = V_{33}V_{88}V_{\sigma\sigma} \nonumber\\ \label{eq: spin pt}
&+ & 2V_{38}V_{3\sigma}V_{8\sigma} - V_{38}^2V_{\sigma\sigma} - V_{3\sigma}^2V_{88} - V_{8\sigma}^2V_{33},\ \ 
\eeq
where each subscript stands for a derivative with respect to the respective variable. The two spinodals merge at the CEP. At this point, the third derivative of the potential along the direction of the eigenvector $\vec{v}_0$ of the Hessian with zero eigenvalue also vanishes
\begin{equation}\label{eq: crit end pt}
   0=\left[\vec{v}_0\cdot\nabla\right]^3V(\sigma,r_3,r_8),
\end{equation}
with $\nabla=(\partial_\sigma,\partial_3,\partial_8)$ and 
\begin{equation}
   \vec{v}_0=(V_{88}V_{\sigma\sigma}-V_{8\sigma}^2,V_{3\sigma}V_{8\sigma}-V_{38}V_{\sigma\sigma},V_{38}V_{8\sigma}-V_{3\sigma}V_{88}).
\end{equation}
Thus, the CEP $(\mu_{\rm CEP},T_{\rm CEP},\sigma_{\rm CEP},r_{3,\rm CEP},r_{8,\rm CEP})$ is found by solving the 5 equations \eqref{eq: saddle pt}, \eqref{eq: spin pt}, \eqref{eq: crit end pt}.

\section{Vacuum quark mass potential}\label{app:VofM}
Consider the quark mass potential in the vacuum ($\smash{T=\mu=0}$):
\beq
V(M)\simeq\frac{(M-m_l)^2}{2G}-\frac{3N_f}{\pi^2}\int_0^\Lambda dq\,q^2\,\varepsilon_q\,.
\eeq
We have
\beq
V'(M)\simeq-\frac{m_l}{G}+M\left[\frac{1}{G}-\frac{3N_f}{\pi^2}\int_0^\Lambda dq\,\frac{q^2}{\varepsilon_q}\right],
\eeq
and
\beq
V''(M) & \simeq & \frac{1}{G}-\frac{3N_f}{\pi^2}\int_0^\Lambda dq\,q^2\left[\frac{1}{\varepsilon_q}-\frac{M^2}{\varepsilon_q^3}\right]\nonumber\\
& = &  \frac{1}{G}-\frac{3N_f}{\pi^2}\int_0^\Lambda  dq\,\frac{q^4}{\varepsilon_q^3}\,.
\eeq
The right-hand side of $V''(M)$ is a monotonically increasing function of $M^2$, varying from
\beq
V''(0)=\frac{1}{G}-\frac{3N_f}{2\pi^2}\Lambda^2\,,
\eeq
to
\beq
V''(\pm\infty)=\frac{1}{G}\,.
\eeq

\begin{figure}[t]
    \centering
    \includegraphics[width=0.9\linewidth]{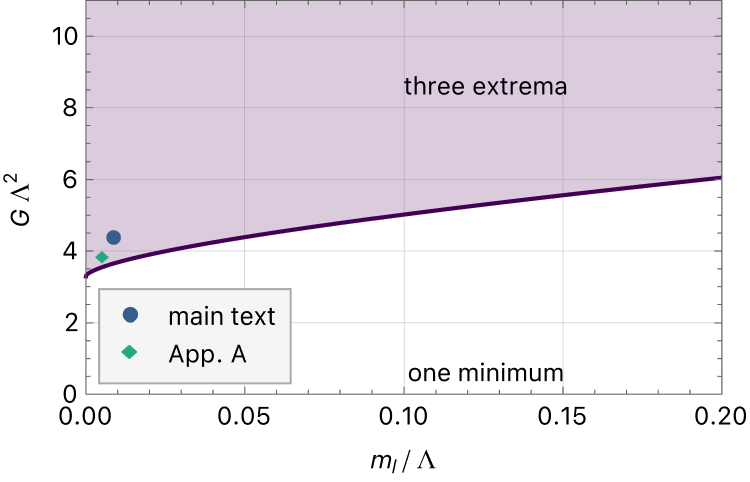}
    \caption{Regions in the plane $(m_l/\Lambda,G\Lambda^2)$ where the vacuum quark mass potential has three extrema, one minimum located in the region $\smash{M>0}$ and another minimum and a maximum located in the region  $\smash{M<0}$. The points represent the parameters considered in this work, either in the main text or in App.~\ref{app: params}.}
    \label{fig:params}
\end{figure}

For our choice of parameters, it is easily checked that $V''(0)$ is negative and thus $V''(M)$ vanishes at some $\smash{M=\pm X}$. More precisely, $V''(M)$ is positive for $\smash{M<-X}$, negative for $-X<M<X$, and again positive for $\smash{M>X}$. It follows that $V'_0(M)$ is monotonically increasing from $-\infty$ to $V'(-X)$ over $\smash{M<-X}$, then monotonically decreasing from $V'(-X)$ to $V'(X)$ over $\smash{-X<M<X}$ and then again monotonically increasing from $V'(X)$ to $+\infty$ over $\smash{M>X}$. Moreover, since $\smash{V'(0)=-m_l/G<0}$, we know that $\smash{V'(X)<0}$ and then that $V'(M)$ has no zero over $\smash{0<M<X}$ and a unique zero over $\smash{M>X}$. By construction, this zero is nothing but $\smash{M=M_0}$. We are left to discuss whether there could be zeros over $\smash{M<0}$. This occurs in case $V'_0(-X)>0$. Let us then evaluate $V'(\pm X)$. We have
\beq
V'(\pm X)=-\frac{m_l}{G}\pm X\left[\frac{1}{G}-\frac{3N_f}{\pi^2}\int_0^\Lambda  dq\,\frac{q^2}{\varepsilon_q}\right],
\eeq
where it is implicitly understood that $\varepsilon_q$ is evaluated for $\smash{M^2=X^2}$. The definition of $\pm X$ is that $\smash{V''(\pm X)=0}$ and thus
\beq
0=\frac{1}{G}-\frac{3N_f}{\pi^2}\int_0^\Lambda  dq\,\frac{q^4}{\varepsilon_q^3}\,,\label{eq:1}
\eeq
where it is again implicitly understood that $\varepsilon_q$ is evaluated for $\smash{M^2=X^2}$. Then
\beq
V'(\pm X)=-\frac{m_l}{G}\mp \frac{3N_f}{\pi^2} X^3\int_0^\Lambda  dq\,\frac{q^2}{\varepsilon_q^3}\,.
\eeq
As announced above, $\smash{V'(X)<0}$ but $\smash{V'(-X)>0}$ imposes a constraint on the parameters:
\beq
m_l<X^3\frac{\int_0^\Lambda dq\,q^2/\varepsilon_q^3}{\int_0^\Lambda dq\,q^4/\varepsilon_q^3}\,.\label{eq:2}
\eeq
where we have made use of Eq.~(\ref{eq:1}). In fact, Eqs.~(\ref{eq:1}) and (\ref{eq:2}) provide a parametric  definition of a region in the $(m_l/\Lambda,G\Lambda^2)$ plane. This region is represented in Fig.~\ref{fig:params}. Note that, from Eq.~(\ref{eq:1}) and because $X\geq 0$, we have
\beq
\frac{1}{G}<\frac{3N_f}{\pi^2}\int_0^\Lambda  dq\,q=\frac{3N_f}{2\pi^2}\Lambda^2\equiv \frac{1}{G_c}\,,
\eeq
and thus $\smash{G\geq G_c}$. The lower bound $G_c$ corresponds to the critical coupling below which there is no spontaneous breaking of chiral symmetry in the chiral limit $\smash{m_l\to 0}$.

Note finally that the potential rewrites as $-m_lM/G$ plus a contribution symmetric under $\smash{M\to -M}$. Since $m_l$ is positive, it follows that the global minimum is always in the region $\smash{M>0}$.

\section{Polyakov loop at NLO}\label{app: NLO_l}
As explained in the main text, the formulas (\ref{eq:l0}) and (\ref{eq:lb0}) correspond to the leading order Polyakov loops in terms of the gauge field expectation value (\ref{eq:vevA}) in the center-symmetric Landau gauge. Even though these expressions can be seen as a change of variables going from $(r_3,r_8)$ to another pair of variables $(\ell,\bar\ell)$, the comparison to lattice data for the Polyakov loop is more consistent if one considers next-to-leading order expressions for the Polyakov loops in the center-symmetric CF model considered in this work. This is because the glue potential $V_{csCF}(r_3,r_8)$  used to determine $r_3$ and $r_8$ is itself determined at next-to-leading order. Let us then sketch the calculation of the Polyakov loops. 

To this purpose, we shall adapt the calculation of Ref.~\cite{Reinosa:2015gxn} where the next-to-leading order Polyakov loops were determined within a self-consistent background Landau gauge. The latter depends on a background
\beq
 g \bar A_\mu^a(x)t^a=T\left(\bar r_3\frac{\lambda_3}{2}+\bar r_8\frac{\lambda_8}{2}\right),\label{eq:bg}
\eeq
which is adapted at each temperature and chemical potential to make sure that $\smash{\bar A_\mu^a=\langle A_\mu^a\rangle}$. The center-symmetric Landau gauge used in the present work also relies on a background of the type (\ref{eq:bg}). This time, however, $(\bar r_3,\bar r_8)$ is fixed to a specific value $(4\pi/3,0)$, such that $\langle A_\mu^a\rangle$ is not necessarily equal to $\bar A_\mu^a$, that is $(r_3,r_8)$ in Eq.~(\ref{eq:vevA}) is not necessarily equal to $(\bar r_3,\bar r_8)$ in Eq.~(\ref{eq:bg}), see Refs.~\cite{vanEgmond:2021jyx, MariSurkau:2024zfb} for more details and discussions on the advantages of this approach, as compared to that based on a self-consistent background.

Beyond these differences, the evaluation of the next-to-leading order Polyakov loops in the center-symmetric Landau gauge follows essentially the same steps as in the self-consistent background Landau gauge. The next-to-leading correction to $\ell$ writes
\beq
\delta\ell(r) & = & -\frac{g^2\beta}{6}\int_0^\beta d\tau\,G^{\kappa\lambda}_{00}(\tau,\vec{0})\nonumber\\
& & \hspace{1.0cm}\times\,{\rm tr}\,\{t^\kappa e^{i(\beta-\tau)Tr^kt^k}t^\lambda e^{i\tau Tr^lt^l}\}\,,\label{eq:trace}
\eeq
where $G^{\kappa\lambda}_{\mu\nu}(\tau,\vec{x})$ is the free propagator in the center-symmetric Landau gauge, with $\kappa$ and $\lambda$ weights of the adjoint representation which serve as a fancy but convenient labeling of color. The adjoint weights arise upon simultaneously diagonalizing the adjoint action of the commuting generators of the algebra, $\smash{[t^j,t^\kappa]=\kappa^j t^\kappa}$. The non-zero adjoint weights are called roots and denoted $\alpha,\dots$ while the vanishing weights are denoted $0^{(j)}$, where the label $j$ is needed because these various zeros are used to label the commuting generators, $\smash{t^{0^{(j)}}=t^j}$. It can be shown that the only non-vanishing components of the free propagator have $\smash{\kappa=\lambda=0^{(j)}}$ or $\smash{\kappa=-\lambda=-\alpha}$.

The trace in Eq.~(\ref{eq:trace}) is evaluated by introducing the defining weights $\rho,\dots$, which diagonalize the defining action of the commuting generators,\footnote{For Polyakov loops in other representations, one would introduce the weights associated to each representation.} $\smash{t^j|\rho\rangle=\rho^j|\rho\rangle}$. We find
\beq
& & {\rm tr}\,\{t^\kappa e^{i(\beta-\tau)Tr^kt^k}t^\lambda e^{i\tau Tr^lt^l}\}\nonumber\\
& & \hspace{1.0cm}=\sum_{\rho,\sigma} e^{i(\beta-\tau)Tr\cdot\sigma+i\tau Tr\cdot\rho}\langle\rho |t^\kappa|\sigma\rangle\langle\sigma|t^\lambda|\rho\rangle\,.
\eeq
There are two cases to be considered. If $\kappa=\lambda=0^{(j)}$, we find
\beq
& & {\rm tr}\,\{t^je^{i(\beta-\tau)Tr^kt^k}t^j e^{i\tau Tr^lt^l}\}\nonumber\\
& & \hspace{1.0cm}=\sum_{\rho,\sigma} e^{i(\beta-\tau)Tr\cdot\sigma+i\tau Tr\cdot\rho}\langle\rho |t^j|\sigma\rangle\langle\sigma|t^j|\rho\rangle\nonumber\\
& & \hspace{1.0cm}=\sum_{\rho,\sigma} \rho^j\sigma^j\,e^{i(\beta-\tau)Tr\cdot\sigma+i\tau Tr\cdot\rho}\langle\rho|\sigma\rangle\langle\sigma|\rho\rangle\nonumber\\
& & \hspace{1.0cm}=\sum_{\rho} (\rho^j)^2\,e^{ir\cdot\rho}\,,
\eeq
where we have used $\smash{\langle\rho|\sigma\rangle=\delta_{\rho\sigma}}$. Since $G^{jj}_{00}$ does not depend on $j$, we can even sum over $j$, and we find
\beq
& & \sum_j{\rm tr}\,\{t^je^{i(\beta-\tau)Tr^kt^k}t^j e^{i\tau Tr^lt^l}\}\nonumber\\
& & \hspace{1.0cm}=\,\frac{1}{2}\left(1-\frac{1}{N_c}\right)\sum_{\rho} e^{ir\cdot\rho}\nonumber\\
& & \hspace{1.0cm}=\,\frac{N_c-1}{2}\ell(r)\,,
\eeq
where $\ell(r)$ denotes the leading order Polyakov loop.  

Next, we consider the case $\kappa=-\alpha$ and $\lambda=\alpha$. We find
\beq
& & {\rm tr}\,\{t^{-\alpha} e^{i(\beta-\tau)Tr^kt^k}t^\alpha e^{i\tau Tr^lt^l}\}\nonumber\\
& & \hspace{1.0cm}=\sum_{\rho,\sigma} e^{i(\beta-\tau)Tr\cdot\sigma+i\tau Tr\cdot\rho}\langle\rho |t^{-\alpha}|\sigma\rangle\langle\sigma|t^\alpha|\rho\rangle\nonumber\\
& & \hspace{1.0cm}=e^{ir\cdot\sigma}\sum_{\rho,\sigma} e^{-i\tau Tr\cdot\alpha}|\langle\sigma|t^\alpha|\rho\rangle|^2\,,
\eeq
where the sum is restricted to $\rho$ and $\sigma$ such that $\sigma-\rho=\alpha$. There is only one such pair for a given $\alpha$. Moreover, since $|\langle\sigma|t^\alpha|\rho\rangle|^2=1/2$, we arrive at
\beq
& & {\rm tr}\,\{t^{-\alpha} e^{i(\beta-\tau)Tr^kt^k}t^\alpha e^{i\tau Tr^lt^l}\}=\frac{1}{2}e^{ir\cdot\sigma}e^{-i\tau Tr\cdot\alpha}\,.
\eeq
All in all, we arrive at
\beq
\delta\ell(r) & = & -\frac{(N_c-1)g^2\beta}{12}\ell(r)\int_0^\beta d\tau\,G^{00}_{00}(\tau,\vec{0})\nonumber\\
& - & \frac{g^2\beta}{12}e^{ir\cdot\sigma}\int_0^\beta d\tau\,G^{(-\alpha)\alpha}_{00}(\tau,\vec{0})\,e^{-i\tau Tr\cdot\alpha}\,,
\eeq
where a sum over $\alpha$ is implied and $\sigma$ is the first weight of $\alpha$. In Fourier space, we find
\beq
\delta\ell(r) & = & -\frac{(N_c-1)g^2\beta}{12}\ell(r)\int_q G^{00}_{00}(0,\vec{q})\nonumber\\
& - & \frac{g^2\beta}{12}\frac{e^{ir\cdot\sigma}-e^{ir\cdot\rho}}{i}\int_Q \frac{G^{(-\alpha)\alpha}_{00}(Q)}{\omega_n+Tr\cdot\alpha}\,,\label{eq:dl}
\eeq
where a sum over $\alpha$ is implied and $\rho$ is the second weight of $\alpha$, and where, according to Ref.~\cite{Reinosa:2015gxn},
\beq
G^{00}_{00}(0,\vec{q}) & = & \frac{1}{q^2+m^2}\,,\\
G^{(-\alpha)\alpha}_{00}(0,\vec{q}) & = & \frac{q^2}{(Q_\alpha\cdot\bar Q_\alpha)^2+m^2\bar Q^2_\alpha}\,.
\eeq
Note that we can multiply the second term of Eq.~(\ref{eq:dl}) by $2$ and sum only over half the roots, or equivalently, if we denote the weights by $\rho_1$, $\rho_2$ and $\rho_3$, sum over the pairs of weights $(\rho_i,\rho_j)$ such that $j>i$, with $\alpha_{ji}=\rho_j-\rho_i$ the corresponding root. 

We finally arrive at
\beq
\delta\ell(r) & = & -\frac{(N_c-1)g^2\beta}{12}\int_q \frac{1}{q^2+m^2}\ell(r)\nonumber\\
& - & \frac{g^2\beta}{6}\sum_{j>i}\frac{e^{ir\cdot\rho_j}-e^{ir\cdot\rho_i}}{i}\nonumber\\
& & \times\,\int_Q \frac{1}{\omega_n+Tr\cdot\alpha_{ji}}\frac{q^2}{(Q_{\alpha_{ji}}\cdot\bar Q_{\alpha_{ji}})^2+m^2\bar Q^2_{\alpha_{ji}}}\,.\nonumber\\
\eeq
A similar expression for the anti-Polyakov loop is obtained upon replacing $r$ by $-r$. After a convenient change of variables under the sum-integral, we arrive at
\beq
\delta\bar\ell(r) & = & 
-\frac{(N_c-1)g^2\beta}{12}\int_q \frac{1}{q^2+m^2}
\bar\ell(r)\nonumber\\
& + & \frac{g^2\beta}{6}\sum_{j>i}\frac{e^{-ir\cdot\rho_j}-e^{-ir\cdot\rho_i}}{i}\nonumber\\
& & \times\,\int_Q \frac{1}{\omega_n+Tr\cdot\alpha_{ji}}\frac{q^2}{(Q_{\alpha_{ji}}\cdot\bar Q_{\alpha_{ji}})^2+m^2\bar Q^2_{\alpha_{ji}}}\,.\nonumber\\
\eeq

Let us now consider the SU(3) case and assume that $\smash{\mu=0}$, in which case $\smash{r_8=\bar r_8=0}$ in which case $\smash{\delta\ell=\delta\bar\ell}$. We also denote $(r_3,\bar r_3)$ more simply as $(r,\bar r)$ and we recall that, eventually $\smash{\bar r=4\pi/3}$. We find that the next-to-leading order Polyakov loop writes as
\beq
\tilde\ell(r) & = & \left[1-\frac{g^2\beta}{6}\int_q \frac{1}{q^2+m^2}\right]\ell(r)-\frac{g^2\beta}{3}\sin(r/2)\nonumber\\
& & \times\,\int_Q \Bigg[\frac{1}{\omega_n^+}\frac{q^2}{(Q_+\cdot\bar Q_+)^2+m^2\bar Q_+^2}+\,(r\to r/2)\Bigg],\nonumber\\
\eeq
with $\smash{\omega_n^+\equiv\omega_n+Tr}$, $\smash{\bar\omega_n^+\equiv\omega_n+T\bar r}$, $\smash{Q_+=(\omega_n^+,q)}$ and $\smash{\bar Q_+=(\bar\omega_n^+,q)}$ and where by $\smash{r\to r/2}$ we mean both $\smash{r\to r/2}$ and $\smash{\bar r\to\bar r/2}$. We next perform the $q$-integrals:
\beq
\int_q \frac{1}{q^2+m^2} & = & \frac{\Gamma((3-d)/2)}{(4\pi)^{(d-1)/2}}(m^2)^{(d-3)/2}\,,
\eeq
and
\beq
& & \int_Q \frac{1}{\omega_n^+}\frac{q^2}{(Q_+\cdot\bar Q_+)^2+m^2\bar Q_+^2}\nonumber\\
& & \hspace{0.2cm}=\,\int_Q \frac{1}{\omega_n^+}\left[\frac{q^2}{q^2+M^2_{+}}-\frac{q^2}{q^2+M^2_{-}}\right]\frac{1}{M^2_{-}-M^2_{+}}\nonumber\\
& & \hspace{0.2cm}=\,\int_Q \frac{1}{\omega_n^+}\left[\frac{M^2_{+}}{q^2+M^2_{+}}-\frac{M^2_{-}}{q^2+M^2_{-}}\right]\frac{1}{M^2_{+}-M^2_{-}}\nonumber\\
& & \hspace{0.2cm}=\frac{\Gamma((3-d)/2)}{(4\pi)^{(d-1)/2}}T\!\sum_n\!\frac{1}{\omega_n^+}\frac{(M^2_{+})^{(d-1)/2}-(M^2_{-})^{(d-1)/2}}{M^2_{+}-M^2_{-}}\,.\nonumber\\
\eeq
So, we finally arrive at
\beq
\tilde\ell(r) & = & \left[1-\frac{g^2\beta}{6}\frac{\Gamma((3-d)/2)}{(4\pi)^{(d-1)/2}}(m^2)^{(d-3)/2}\right]\ell(r)\nonumber\\
& - & \frac{g^2}{3}\frac{\Gamma((3-d)/2)}{(4\pi)^{(d-1)/2}}\sin(r/2)\nonumber\\
& \times & \!\sum_n\!\left[\frac{1}{\omega_n^+}\frac{(M^2_{+})^{(d-1)/2}-(M^2_{-})^{(d-1)/2}}{M^2_{+}-M^2_{-}}\!+\!(r\to r/2)\right].\nonumber\\
\eeq
While $d$ can be safely taken equal to $4$ in the first line, the same is not true for the second line as it contains a UV divergence. The latter is extracted by writing $\omega_n^+=\bar\omega_n^++T\Delta r$ with $\smash{\Delta r\equiv r-\bar r}$ and expanding the summand for large $|\bar\omega_n^+|$. One finds
\beq
& & \frac{1}{\omega_n^+}\frac{(M^2_{+})^{(d-1)/2}-(M^2_{-})^{(d-1)/2}}{M^2_{+}-M^2_{-}}\nonumber\\
& & \hspace{0.5cm}=\,\frac{d-1}{2}|\bar\omega_n|^{d-1}\left[\frac{1}{\bar\omega_n^3}+\frac{d-5}{2}\frac{T\Delta r}{\bar\omega_n^4}+\dots\right].
\eeq
We now add and subtract this asymptotic behavior from the summand. In the subtracted terms, we can replace $d$ by $4$, while in the added terms, we need to keep $d$ arbitrary and compute the corresponding sums explicitly (note that, in the corresponding term, $g^2$ needs to be replaced by $g^2\Lambda^{2\epsilon}$). This can be done using the Hurwitz zeta function, see Ref.~\cite{MariSurkau:2024zfb}. We obtain
\begin{widetext}
\beq
\tilde\ell(r) & = & \left[1+\frac{g^2\beta m}{24\pi}\right]\ell(r)-\frac{g^2\beta}{3}\frac{\Gamma((3-d)/2)}{(4\pi)^{(d-1)/2}}\frac{d-1}{2}\sin(r/2)\Lambda^{2\epsilon}T\sum_n \left[|\bar\omega_n|^{d-1}\left(\frac{1}{\bar\omega_n^3}+\frac{d-5}{2}\frac{T\Delta r}{\bar\omega_n^4}\right)+(r\to r/2)\right]\nonumber\\
& + & \frac{g^2}{12\pi}\sin(r/2)\sum_n \left[\frac{1}{\omega_n^+}\frac{(M^2_{+})^{3/2}-(M^2_{-})^{3/2}}{M^2_{+}-M^2_{-}}-\frac{3}{2}|\bar\omega_n|^3\left(\frac{1}{\bar\omega_n^3}-\frac{T\Delta r}{2\bar\omega_n^4}\right)+(r\to r/2)\right].
\eeq
We next use that (see Ref.~\cite{MariSurkau:2024zfb})
\beq
& & \Lambda^{2\epsilon}\sum_n \frac{|\bar\omega_n|^{d-1}}{\bar\omega_n^3}=1-2\left\{\frac{\bar r}{2\pi}\right\}+{\cal O}(\epsilon)\,,\\
& & \Lambda^{2\epsilon}\sum_n \frac{|\bar\omega_n|^{d-1}}{\bar\omega_n^4}=\frac{1}{2\pi T}\Bigg[\frac{1}{\epsilon}+\ln\frac{\Lambda^2}{4\pi^2 T^2}
-\,\psi\left(\left\{\frac{\bar r}{2\pi}\right\}\right)-\psi\left(1-\left\{\frac{\bar r}{2\pi}\right\}\right)+{\cal O}(\epsilon)\Bigg]\,,
\eeq
where $\psi$ denotes the digamma function. We finally arrive at
\beq
\tilde\ell(r) & = & \left[1+\frac{g^2\beta m}{24\pi}\right]\ell(r)-\frac{3g^2}{64\pi^2}\Delta r \sin(r/2)\left[\frac{1}{\epsilon}+\frac{10}{3}+\ln\frac{\bar\Lambda^2}{16\pi^2 T^2}\right]+\frac{g^2}{4\pi}\sin(r/2)\left[1-\left\{\frac{\bar r}{2\pi}\right\}-\left\{\frac{\bar r}{4\pi}\right\}\right]\nonumber\\
& + & \frac{g^2}{32\pi^2}\Delta r\sin(r/2)\left[\psi\left(\left\{\frac{\bar r}{2\pi}\right\}\right)+\psi\left(1-\left\{\frac{\bar r}{2\pi}\right\}\right)+\frac{1}{2}\psi\left(\left\{\frac{\bar r}{4\pi}\right\}\right)+\frac{1}{2}\psi\left(1-\left\{\frac{\bar r}{4\pi}\right\}\right)\right]\nonumber\\
& + & \frac{g^2}{12\pi}\sin(r/2)\sum_n \left[\frac{1}{\omega_n^+}\frac{(M^2_{+})^{3/2}-(M^2_{-})^{3/2}}{M^2_{+}-M^2_{-}}-\frac{3}{2}|\bar\omega_n|^3\left(\frac{1}{\bar\omega_n^3}-\frac{T\Delta r}{2\bar\omega_n^4}\right)+(r\to r/2)\right].
\eeq
\end{widetext}
The presence of a divergence is expected because, unlike $\bar r$ which is not renormalized \cite{MariSurkau:2024zfb}, the difference $r-\bar r$ gets rescaled by a renormalization factor $Z=Z^{1/2}_aZ_g$. After the rescaling, the leading order Polyakov loop becomes
\beq
\ell(r) & = & \frac{1}{3}\sum_\rho e^{ir\cdot\rho}\nonumber\\
& \to & \frac{1}{3}\sum_\rho e^{i(\bar r+Z\Delta r)\cdot\rho}\nonumber\\
& \to & \frac{1}{3}\sum_\rho e^{i(r+\delta Z\Delta r)\cdot\rho}\nonumber\\
& \to &\ell(r)+i\frac{\delta Z}{3}\Delta r\cdot\sum_\rho  \rho\,e^{ir\cdot\rho}\nonumber\\
& \to &\ell(r)-\frac{\delta Z}{3}\Delta r\sin(r/2)\,
\eeq
with
\beq
\delta Z=-\frac{9g^2}{64\pi^2}\frac{1}{\epsilon}+z_f\,.\label{eq:dZ}
\eeq
and where we have used that $\smash{r_8=0}$ in the last step. Plugging this in the expression above, we arrive at
\begin{widetext}
\beq
\tilde\ell(r) & = & \left[1+\frac{g^2\beta m}{24\pi}\right]\ell(r)-\frac{1}{3}\Delta r \sin(r/2)\left\{\delta Z+\frac{9g^2}{64\pi^2}\left[\frac{1}{\epsilon}+\frac{10}{3}+\ln\frac{\bar\Lambda^2}{16\pi^2 T^2}\right]\right\}+\frac{g^2}{4\pi}\sin(r/2)\left[1-\left\{\frac{\bar r}{2\pi}\right\}-\left\{\frac{\bar r}{4\pi}\right\}\right]\nonumber\\
& + & \frac{g^2}{32\pi^2}\Delta r\sin(r/2)\left[\psi\left(\left\{\frac{\bar r}{2\pi}\right\}\right)+\psi\left(1-\left\{\frac{\bar r}{2\pi}\right\}\right)+\frac{1}{2}\psi\left(\left\{\frac{\bar r}{4\pi}\right\}\right)+\frac{1}{2}\psi\left(1-\left\{\frac{\bar r}{4\pi}\right\}\right)\right]\nonumber\\
& + & \frac{g^2}{12\pi}\sin(r/2)\sum_n \left[\frac{1}{\omega_n^+}\frac{(M^2_{+})^{3/2}-(M^2_{-})^{3/2}}{M^2_{+}-M^2_{-}}-\frac{3}{2}|\bar\omega_n|^3\left(\frac{1}{\bar\omega_n^3}-\frac{T\Delta r}{2\bar\omega_n^4}\right)+(r\to r/2)\right].
\eeq
We check that the known value for $\delta Z$ precisely cancels the divergence. We finally arrive at the finite expression
\beq
\tilde\ell(r) & = & \left[1+\frac{g^2\beta m}{24\pi}\right]\ell(r)-\frac{1}{3}\Delta r \sin(r/2)\left\{z_f+\frac{9g^2}{64\pi^2}\left[\frac{10}{3}+\ln\frac{\bar\Lambda^2}{16\pi^2 T^2}\right]\right\}+\frac{g^2}{4\pi}\sin(r/2)\left[1-\left\{\frac{\bar r}{2\pi}\right\}-\left\{\frac{\bar r}{4\pi}\right\}\right]\nonumber\\
& + & \frac{g^2}{32\pi^2}\Delta r\sin(r/2)\left[\psi\left(\left\{\frac{\bar r}{2\pi}\right\}\right)+\psi\left(1-\left\{\frac{\bar r}{2\pi}\right\}\right)+\frac{1}{2}\psi\left(\left\{\frac{\bar r}{4\pi}\right\}\right)+\frac{1}{2}\psi\left(1-\left\{\frac{\bar r}{4\pi}\right\}\right)\right]\nonumber\\
& + & \frac{g^2}{12\pi}\sin(r/2)\sum_n \left[\frac{1}{\omega_n^+}\frac{(M^2_{+})^{3/2}-(M^2_{-})^{3/2}}{M^2_{+}-M^2_{-}}-\frac{3}{2}|\bar\omega_n|^3\left(\frac{1}{\bar\omega_n^3}-\frac{T\Delta r}{2\bar\omega_n^4}\right)+(r\to r/2)\right].
\eeq
Note finally that $\delta Z$ is nothing but the ghost renormalization factor $\delta Z_c$ in any scheme using the Taylor condition. Then $z_f$ is the same in both the vanishing momentum and IR-safe scheme used in \cite{MariSurkau:2024zfb}:
\beq
z_f & = & -\frac{3g^2}{64\pi^2}\left[3\ln\frac{\bar\Lambda^2}{m^2}+5+\frac{m^2}{\mu^2} + \frac{\mu^2}{m^2}\ln\frac{\mu^2}{m^2}-\frac{(\mu^2+m^2)^3}{\mu^4m^2}\ln\left(1+\frac{\mu^2}{m^2}\right)\right].
\eeq
For $\smash{\bar r=4\pi/3}$, we finally arrive at
\beq
\tilde\ell(r) & = & \left[1+\frac{g^2\beta m}{24\pi}\right]\ell(r)-\frac{3g^2}{64\pi^2}\Delta r \sin(r/2)\left[\frac{5}{3}+\ln\frac{27e^{2\gamma}m^2}{16\pi^2 T^2}-\frac{m^2}{3\mu^2}-\frac{\mu^2}{3m^2}\ln\frac{\mu^2}{m^2}+\frac{(\mu^2+m^2)^3}{3\mu^4m^2}\ln\left(1+\frac{\mu^2}{m^2}\right)\right]\nonumber\\
& + & \frac{g^2}{12\pi}\sin(r/2)\sum_n \left[\frac{1}{\omega_n^+}\frac{(M^2_{+})^{3/2}-(M^2_{-})^{3/2}}{M^2_{+}-M^2_{-}}-\frac{3}{2}|\bar\omega_n|^3\left(\frac{1}{\bar\omega_n^3}-\frac{T\Delta r}{2\bar\omega_n^4}\right)+(r\to r/2)\right]. \label{eq:NLO_PolyL}
\eeq
\end{widetext}
This is the expression that was used in producing the next-to-leading order curve in Fig.~\ref{fig: ord par vs lat dat}. We note that the above derivation is not valid at finite $\mu$ where ${r_8\neq0}$ and Fig.~\ref{fig: ord par vs lat dat} only shows ${\mu=0}$ results, but since we mostly find ${r_8\ll1}$ employing it anyways should provide a decent estimate of the next-to-leading order contributions even away from ${\mu=0}$. A full derivation with ${r_8\neq0}$ will be reported elsewhere.

The perturbative calculation of the Polyakov loop can naturally be reorganized as a calculation of the exponential of the perturbatively calculated free energy change induced by the static quark, see e.g.~\cite{Berwein:2015ayt} and references therein for details. This becomes particularly relevant at next-to-next-to-leading order, where the higher-order terms get rearranged together with squared NLO terms. In standard gauges, at NLO ($\mathcal{O}(g^2)$) the exponentiation gives simply ${\Tilde{\ell}\simeq1+\delta\ell\to\exp(\delta\ell)}$. However, with the non-trivial gauge background the NLO Polyakov loop can instead be calculated as
\begin{equation}
    \Tilde{\ell}(r) \simeq \ell(r)\exp\left(\frac{\delta\ell(r)}{\ell(r)}\right),
\end{equation}
which can be a more substantial reorganization, especially when $\ell(r)$ is small, affecting the temperature dependence of the Polyakov loop. Computing the NLO Polyakov loop in this manner instead of \eqref{eq:NLO_PolyL}, we find that with an appropriate constant rescaling it can be made to follow the lattice data near the transition slightly better, but is slightly worse at higher and lower temperatures. Higher-order corrections may be necessary to further improve the behavior. Note that we  have also not included an HTL resummation. We do not show the results to avoid cluttering Fig.~\ref{fig: ord par vs lat dat}.

\section{Inflection points}\label{app:inflections}
In the main text, we have discussed the utility of following the inflection points, in the $(\mu,T)$ plane, of the net quark number responses $\Delta Q_q$ and $\Delta Q_{\bar q}$ as probes of the phase diagram, which can reveal connections between various singular points, and also of the chiral condensate and Polyakov loop between the CEP and the $T$-axis, which are commonly used to define the crossover temperatures. As we alluded to above, these quantities can have other inflection points, which should not necessarily be interpreted as corresponding to phase crossovers.

This can be understood clearly for the chiral condensate by looking at Fig.~\ref{fig: ord par} for the largest ${\mu\simeq372\,}$MeV curve. It has an inflection as it evolves from the very small ${T=0}$ value of about $11\%$ of the vacuum value to the high $T$ limit ${\sim-m_l}$. Reasonably, both of those limits can be seen as corresponding to an approximately chirally restored symmetry, especially compared to the ten times larger vacuum or low $T$ and ${\mu<M_0}$ condensate. Then, the inflection does not correspond to a chiral transition in the same sense as at low $\mu$. The changes of the current quark masses may still be physical in the sense that at higher energy scales they decrease (run) towards the asymptotically free, massless limits, but it is questionable how much of that is captured by the current, non-renormalizable NJL model. It would rather be interesting to extend the model to include diquark degrees of freedom, in which case a real phase transition or crossover should appear in the high $\mu$ regime \cite{Son:1998uk, Roessner:2006xn, Braun:2019aow}.

Similar arguments apply to the ${\mu\simeq372\,}$MeV Polyakov loops in Fig.~\ref{fig: ord par}, which clearly already signal deconfinement at ${T=0}$, they just happen to have another inflection as they asymptotically approach 1. Here, the behavior depends more strongly on the employed Polyakov loop potential, but as we argued in Sec.~\ref{sec:T0}, irrespective of the used potential, above ${\mu=M_0}$ quarks should be considered as deconfined in the PNJL model. Whether the inflections can be interpreted as a change from a quarkyonic regime to a pure quark (and gluon) phase is then unclear.

\begin{figure}[ht]
    \centering
    \includegraphics[width=0.9\linewidth]{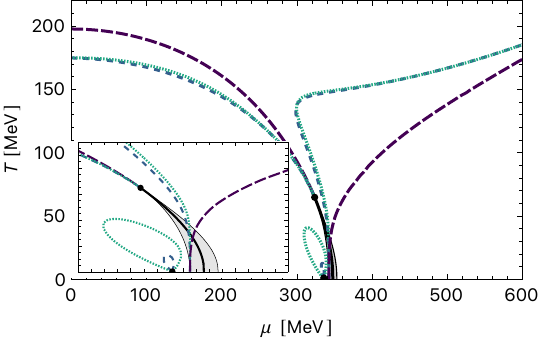}
    \caption{Inflection points of the usual (de)confinement and chiral order parameters, showing the additional inflections in the already deconfined and chirally restored phase and around the density onset point at ${\mu=M_0}$. The long dashed, dark lines corresponds to $\sigma$, dashed to $\ell$, and dotted, light to $\bar\ell$. The inset shows a zoom on the region around the spinodals.}
    \label{fig: inflections}
\end{figure}

In Fig.~\ref{fig: inflections}, for completeness, we now show all the inflection points of the Polyakov loops $\ell$ and $\bar\ell$, and the chiral condensate $\sigma$, along with the spinodal region and first order jump.\footnote{Note that we use the parameters of the main text again, not of App.~\ref{app: params}.} Interestingly, the Polyakov loops are sensitive to the point ${\mu=M_0}$, as also visible in Fig.~\ref{fig: ord pars at mu=M}. However, their inflections form a closed loop originating and ending at the point $(M_0,0)$, and it is hard to assign a phase change to them. Moreover, they do not connect this point to the CEP, as the net quark number responses $\Delta Q_{q,\bar q}$ do. The bending of the Polyakov loops is also stronger for $\bar\ell$ than for $\ell$ at positive $\mu$, as can be seen in Fig.~\ref{fig: ord pars at mu=M}. 

At values of $\mu$ to the right of the usual chiral/deconfinement transition, both the Polyakov loops and the chiral condensate have another set of inflections, see Fig.~\ref{fig: inflections}. In fact, these inflections continue on the same solution branch into the metastable region, to the left of the first-order jump. They then disappear towards $T\to0$ at the edge where that solution branch disappears. No further inflection points appear in the solution branch corresponding to the chirally broken phase.

\bibliography{references}
\end{document}